# Correlating 3D atomic defects and electronic properties of 2D materials with picometer precision

Xuezeng Tian[1†], Dennis S. Kim[1†], Shize Yang[2†‡], Christopher J. Ciccarino[3,4], Yongji Gong[5], Yongsoo Yang[1§], Yao Yang[1], Blake Duschatko[3], Yakun Yuan[1], Pulickel M. Ajayan[5], Juan-Carlos Idrobo[2], Prineha Narang[3] & Jianwei Miao[1*]

[1]*Department of Physics & Astronomy and California NanoSystems Institute, University of California, Los Angeles, CA 90095, USA.* [2]*Center for Nanophase Materials Sciences, Oak Ridge National Laboratory, Oak Ridge, Tennessee 37831, USA.* [3]*John A. Paulson School of Engineering and Applied Sciences, Harvard University, Cambridge, MA, USA.* [4]*Department of Chemistry and Chemical Biology, Harvard University, Cambridge, MA, USA.* [5]*Department of Mechanical Engineering & Materials Science, Rice University, Houston, Texas 77005, USA.* [†]*These authors contributed equally to this work.* [‡]*Current address: Center for Functional Nanomaterials, Brookhaven National Laboratory, Upton, New York 11973, USA.* [§]*Current address: Department of Physics, Korea Advanced Institute of Science and Technology, Daejeon 34141, Korea.* [*]*e-mail: miao@physics.ucla.edu*

**The exceptional electronic, optical and chemical properties of two-dimensional materials strongly depend on the 3D atomic structure and crystal defects. Using Re-doped MoS₂ as a model, here we develop scanning atomic electron tomography (sAET) to determine the 3D atomic positions and crystal defects such as dopants, vacancies and ripples with a precision down to 4 picometers. We measure the 3D bond distortion and local strain tensor induced by single dopants for the first time. By directly providing experimental 3D atomic coordinates to density functional theory (DFT), we obtain more truthful electronic band structures than those derived from conventional DFT calculations relying on relaxed 3D atomic models, which is confirmed by photoluminescence measurements.**

**We anticipate that sAET is not only generally applicable to the determination of the 3D atomic coordinates of 2D materials, heterostructures and thin films, but also could transform *ab initio* calculations by using experimental 3D atomic coordinates as direct input to better predict and discover new physical, chemical and electronic properties.**

Due to the reduced dimensionality, the properties and functionality of 2D layered materials are strongly influenced by atomic defects such as dopants, vacancies, dislocations, grain boundaries, strains, ripples and interfaces[1-11]. Although aberration-corrected electron microscopy and scanning probe microscopy can image these materials at atomic resolution, they only provide 2D projection images or surface atomic structures[1,7-13]. Atomic electron tomography (AET) allows the determination of 3D atomic structure of crystal defects and chemical disorder systems[14-20], and has recently been advanced to observe crystal nucleation at 4D atomic resolution[21]. However, AET has thus far been limited to metallic nanoparticles and needle-shaped samples. Application of AET to 2D materials and van der Waals heterostructures would open a new frontier in 3D atomic structure characterization, but requires overcoming three obstacles. First, these materials are electron beam sensitive and the experiments must be performed with low electron doses[1,2]. Second, due to the geometric constraint of 2D materials and heterostructures and their suspended nature, the tilt range of data acquisition is limited in aberration-corrected electron microscopy[22]. Third, the 3D precision of experimental atomic coordinates must be on the picometer scale so that they can be used as direct input to quantum mechanical calculations to determine physical, material and electronic properties[23-25]. Here, we developed sAET to overcome these limitations and determine the 3D coordinates of individual atoms in Re-doped $MoS_2$ monolayers with picometer precision. We identified 3D crystal defects such as dopants, vacancies and atomic-scale ripples and measured 3D bond distortion and local strain tensor induced by single Re dopants. We showed that the experimental 3D atomic coordinates can be used as direct input to DFT calculations to derive more truthful electronic band structures than those obtained from conventional DFT with relaxed atomic coordinates.

**Scanning atomic electron tomography (sAET)**

Doping in semiconductors has been an important tool for tailoring their chemical, optical and electronic properties. However, due to the reduced dimensionality, doping in 2D materials is beyond a simple substitutional electron donor or acceptor. Taking Mo based transition metal dichalcogenides (TMD) for example, Re dopants have been identified as a good candidate for n-type doping[26,27]. A high concentration of Re dopants in $MoS_2$ monolayers led to the phase transition from a trigonal prismatic (1H) structure to a distorted octahedral (1T') structure, which increases the catalytic activities[28]. But a fundamental understanding on how the early stage phase transition takes place at the atomic scale remains unknown. To shed light on how single dopants in 2D materials perturb their 3D local structure and induce local strain tensor, we conducted sAET experiments using an aberration-corrected scanning transmission electron microscope (STEM), operated at 60 kV with an annular dark-field mode (ADF) detector. Two data sets were collected from the Re-doped $MoS_2$ monolayer, each consisting of 13 projections with small double tilt ranges (Supplementary Fig. 1 and Table 1). A low-dose data acquisition scheme was implemented with a total dose of $4.1 \times 10^5$ e/$Å^2$ for each data set (Methods), which is comparable to that of a single high-resolution ADF-STEM image[28]. Comparison of the first and last images of each data set indicates the consistency of the atomic structure during data acquisition (Supplementary Fig. 2).

After processing the two data sets (Methods, Supplementary Figs. 3 and 4), we performed sAET reconstruction of the aligned projections (Fig. 1a). sAET takes advantage of the relationship, $d \propto \frac{D}{N}$, where $d$ is the 3D resolution, $D$ is the size of the object to be reconstructed and $N$ is the number of projections[29]. For a small $N$ limited by the electron dose, we can improve $d$ by reducing $D$. To implement sAET, we first chose a 3D window of 60×60×60 voxels and located the corresponding region in each of the 13 projections (Fig. 1a). We then scanned the window along the x and y axis with a step size of 30 voxels. At each step, we identified the corresponding regions in the projections for each window. After all the projections were partitioned into small image stacks, each 3D window was reconstructed by the generalized Fourier iterative reconstruction (GENFIRE) algorithm[30]. All the 3D windows were then stitched together to form a full reconstruction, where the overlap volume between neighboring windows containing artifacts due to boundary effects was discarded (Methods). Supplementary Figs. 5, 6 and Video 1 show the reconstructions of data sets 1 and 2, where all the Re dopants are direct replacements of Mo atoms. The side view images in Supplementary Fig. 5a (inset) illustrate the intensities of Mo, S, and Re atoms with S vacancies indicated by arrows. From each reconstruction, the 3D atomic coordinates and chemical species were traced, refined and corroborated (Methods). Data set 1 consists of 1381 S, 686 Mo, 21 Re atoms and 15 S vacancies (Supplementary Fig. 5b and Video 2), while data set 2 comprises 1083 S, 531Mo, 16 Re atoms and 4 S vacancies (Supplementary Fig. 6b).

To estimate the 3D precision of the atomic positions, we implemented two independent methods for cross validation. First, we identified the x and y coordinates of the Re and Mo atoms in each projection of the data sets using a Gaussian fitting (Supplementary Fig. 7a), whereas the positions of individual S atoms could not be accurately localized from an individual 2D projection due to their weak contrast in ADF images. From these 2D coordinates of Re and Mo atoms, we used least square fitting to calibrate the tilt angles (Supplementary Table 2) and determine their 3D atomic coordinates simultaneously (Supplementary Fig. 7b). The Re and Mo atomic coordinates are consistent with those obtained by sAET with a root mean square deviation (RMSD) of 13 pm and 2 pm, respectively (Supplementary Figs. 7c and d). Second, we performed multislice simulations to calculate 13 projections from the experimental 3D atomic model (Methods). The multislice projections are in good agreement with the corresponding experimental projections (Supplementary Fig. 8). By applying the same reconstruction, atom tracing, and refinement procedures, we obtained a new 3D atomic model from the 13 multislice projections with all atoms and defects correctly identified. The RMSD between the experimental and new atomic models is 15 pm, 12 pm and 4 pm for S, Re and Mo atoms, respectively (Fig. 1b). The 3D precision of the Re and Mo atoms is consistent with those obtained by the least square fitting method.

To show the advantage of sAET over AET for the 3D reconstruction of 2D layered materials, we reconstructed the experimental Re-doped $MoS_2$ data sets using AET (Fig. 1c). Due to the small tilt range and the limited number of projections, all atoms in the AET reconstruction are elongated along the z-axis (the beam direction). This atom elongation problem is significantly alleviated by sAET (Fig. 1c), allowing more precise determination of the 3D coordinates of individual atoms. To demonstrate the applicability of sAET to general 2D materials, we performed numerical simulations on the 3D reconstruction of a $MoSe_2$-$WSe_2$ heterostructure with a $5°$ rotational mismatch between the top and bottom layers. This type of heterostructures has recently received considerable attention due to its moiré exciton properties[31,32]. Using multislice simulations, we calculated a tilt series of 15 projections from the heterostructure consisting of atomic defects (Methods). From the same data, we reconstructed and traced the 3D atomic positions using sAET and AET. Supplementary Fig. 9 shows the sAET reconstruction and the traced atomic positions in six atom layers, which are in good agreement the original 3D atomic model. The RMSD between the sAET reconstruction and the original atomic model is 8 pm, 5 pm and 13 pm for Mo, W and Se atoms, respectively. But the atomic shapes in the AET reconstruction are significantly elongated along the z-axis and are barely traceable (Fig. 1d). These results show the power of sAET to determine the 3D atomic positions and crystal defects in general 2D materials with picometer precision.

**Measuring the 3D bond distortion and local strain tensor induced by single dopants**

From the experimental 3D atomic coordinates, we observed atomic-scale ripples and estimated the standard deviation of the Mo and Re atoms along the z-axis to be 19 pm and 16 pm for data sets 1 and 2, respectively (Fig. 2a, b, Supplementary Fig. 6c and d). The magnitude and wavelength of the ripples in Re-doped $MoS_2$ are within the range of previously reported values[33-39]. Compared to other techniques to characterize ripples in 2D materials[33-39], sAET is a real-space method that can reveal 3D atomic-scale ripples with picometer precision. Such a high precision allowed us to measure the 3D bond distortion induced by single dopants. We found the Re-Re pair and the Re-S vacancy pair induce larger distortion than single Re dopants (Fig. 2c). We determined the Mo-S and Re-S bond lengths to be 2.39±0.17 Å and 2.20±0.29 Å, respectively (Fig. 2d). Statistically, the Re-S bond is 19 pm shorter than the Mo-S bond. With S vacancies, the opposite Re/Mo-S bond length was reduced to 2.02±0.28 Å (Fig. 2d). Although there are only 37 Re dopants in the two data sets, we found that 10 out of a total of 19 S vacancies are situated next to the Re dopants, which is due to the lower vacancy formation energy of neighbouring Re dopants[26-28]. Furthermore, we also determined the deviation of the bond angles from the 1H structure. The angles between the Re-S bonds in the same and opposite atomic layer are 85.1±9.4° and 78.5±8.7°, respectively, whereas the corresponding angles of the Mo-S bonds are 82.6±6.6° and 80.6±4.8° (Fig. 2e).

Next, we measured the 3D atomic displacements and full strain tensors of the 2D material. Figure 3a-c shows the displacements of the Mo and Re atoms in data set 1 along the x-, y- and z-axis, respectively. From the atomic displacements, we determined the full 3D strain tensor of both data sets. Figure 3d-i and Supplementary Fig. 10 show the six components of the strain tensor in three atomic layers, where we observed both the non-local and local strain tensor. The non-local strain was induced by the flexible and thin characteristics of the 2D material suspended on a TEM grid. To decouple the non-local and local strain tensor, we chose spheres with a radius of 3.16 Å centered at Re and Mo atoms in the two data sets. We measured the local strain tensor within the spheres and observed that the average local strains of $\varepsilon_{xx}$, $\varepsilon_{yy}$ and $\varepsilon_{zz}$ between the Re and Mo atomic positions changed by 0.76%, 1.1% and 1.88%, respectively, while the change of the local strain of $\varepsilon_{xy}$, $\varepsilon_{xz}$ and $\varepsilon_{yz}$ is very small (Fig. 4).

These results enable us to understand the dopant induced phase transition in $MoS_2$ at the single-atom level. Previous studies have reported the phase transition of $MoS_2$ from the 1H structure to the 1T' structure with a high concentration of Re dopants[26-28]. However, how the early stage phase transition initiates at the single dopant remains unknown. According to crystal field theory[40], the electronic structure and preferred phase of TMDs strongly depends on the coordination environment of the transition metal and the $d$-electrons[40,41]. As a group 6 TMD, $MoS_2$ has 2 non-bonding $d$-electrons ($d^2$), which prefer to fill in the $d_{z2}$ orbital of the 1H phase in ambient conditions. But Re has 3 non-bonding $d$-electrons ($d^3$) and tends to form an octahedral phase in a TMD[40]. The extra $d$-electrons coming from Re dopants in $MoS_2$ fill higher energy levels of the 1H phase ($d_{xy}$ and $d_{x2-y2}$ orbitals), which destabilizes and distorts the 1H phase towards an octahedral phase[28, 41-44]. Thus our experimental observations provide a new perspective to understand dopant induced phase transition at the single-atom level. It is expected the local distortion created by Re dopants to have a strong consequence on the electronic properties of the 2D material.

**Correlating 3D crystal defects with electronic structures at the single-atom level**

To correlate 3D crystal defects with the electronic structure of the Re-doped $MoS_2$ monolayer, we selected four supercells in different regions of data set 1, each consisting of 6×6×1 unit cells (Fig. 5a-e). The supercells include a dopant-free $MoS_2$ structure, a single Re dopant, double Re dopants, and mixture of a Re dopant and a S vacancy (Fig. 5e). The experimental 3D atomic coordinates of each supercell were used as direct input to DFT to determine the effective band structure of each supercell and revel the underlying electronic symmetries via band unfolding techniques (Methods). For comparison, the same experimental atomic coordinates were relaxed to obtain the equilibrium band structures by DFT (Fig. 5a-d middle). The band structures obtained directly from the experimental coordinates show highly distorted indirect band gaps of the 2D material with a large number of shadow bands. In contrast, the band structures of the relaxed experimental coordinates exhibit direct band gaps with significantly reduced shadow bands[44]. This behavior is consistent throughout the Brillouin zone as seen in each supercell density of states (DOS) shown to the right of each band structure calculations (Fig. 5a-d).

Strain is known to strongly influence the band structure of 2D materials[6,11]. To study the effects of the local strain on the band structures, we calculated the principal strains by solving for the eigenvalues and eigenvectors of the strain tensor within each supercell. The principal strains of Fig. 5a-d are [0.4%, -0.23%, -1.15%], [0.5% -1.00%, -0.60%], [0.14%, -0.71%, -1.58%] and [-0.04%, -1.84%, 1.35%], respectively (Methods, Supplementary Fig. 11), indicating that the increase of the local strain breaks the local symmetry and generates more shadow bands. In addition to shadow bands, we also observed defect states in the band structure of the supercell containing one Re dopant (Fig. 5b left and middle), whereas more defect states appear between the band gap in the supercell having two Re dopants (Fig. 5c left and middle). The striking differences between the band structures of the experimental and relaxed atomic coordinates suggest that the experimental structure is in a metastable state due to the complex strain distribution induced by 3D crystal defects and the interactions with the rest of the lattice, while the relaxed structure is in the global equilibrium state. The metastable structure is feasible as Re dopants in $MoS_2$ produce defect wave functions that delocalize with slowly decaying tails[45]. This delocalization of defect states with increasing concentration in various clusters of dopants results in a metastable structure, which clearly exhibits different electronic band structures (Fig. 5a-d). To quantitatively analyze the band structure, we averaged the DOS of the experimental structures in the four supercells (Fig. 5f, black curve), where the defect states and shadow bands from different energies and regions fill in the band gap. As a reference, the DOS of the relaxed $MoS_2$ structure is shown in Fig. 5f (blue curve). Although the DOS was obtained from four supercells, based on the strain maps we expect a similar result if the whole sample was used for DOS calculations.

To experimentally study the effect of the band structure, we measured photoluminescence (PL) and Raman spectra of Re-doped $MoS_2$ and pristine $MoS_2$ monolayers synthesized under the same conditions (Methods, Fig. 5g and Supplementary Figs. 12 and 13). After examining several Re-doped $MoS_2$ monolayer samples, we found most of the PL spectra were completely quenched with no visible peak above the background. Figure 5g (black curve) shows the PL spectrum of one of the samples with a very weak exciton peak. The scale of the Re-doped $MoS_2$ spectrum relative to the background and pristine $MoS_2$ is shown in Supplementary Fig. 12. Compared to the pristine $MoS_2$ data (Fig. 5g, blue curve), the Re-doped $MoS_2$ spectrum exhibits much lower peak intensity with the peak position only slightly shifted towards lower energies. This observation agrees with our effective band structure calculations (Fig. 5f). Each region in Fig. 5 create defect states within the gap at different energies, essentially creating a metal-like DOS. But the density within the gap is relatively low and a semiconductor band gap can be realized. From the DOS calculations, we expect only up to a small feature visible in the photoluminescence spectrum as the system is a highly distorted indirect band gap semiconductor or metal-like behavior (Fig. 5f). This is consistent with what we observed experimentally as certain regions show a small peak (Fig. 5g) and others quenched spectra. Supplementary Fig. 13 shows the Raman spectra of the Re-doped $MoS_2$ and pristine $MoS_2$ monolayers. For both the in-plane and out-of-plane phonon modes, the peak intensity measured from Re-doped $MoS_2$ is about 1/3 of that from pristine $MoS_2$ (Supplementary Fig.

13). The peak intensity reduction in Re-doped $MoS_2$ indicates that the long-range order of the 1H structure of $MoS_2$ was highly distorted by Re dopants, which is confirmed by our direct experimental observations in three dimensions (Figs. 2-4).

Our results demonstrate that doping in 2D materials depends not only on the concentration, but also on the local bonding characteristics and the equilibrium phase of the dopants. For example, $WS_2$ with a few percent Nb dopants shows a p-type direct band gap semiconductor with a strong PL peak[46], whereas both $NbS_2$ and $WS_2$ have the 1H structure. But $MoS_2$ with a similar Re dopant concentration exhibits a quenched PL spectrum (Fig. 5f), whereas $MoS_2$ has the 1H structure and $ReS_2$ has the 1T' structure. The difference in local bonding and equilibrium phase of the dopants appears to account for these stark differences in doping effects at similar concentrations. Furthermore, at a sufficiently lower dopant concentration (0.3%), Re-doped $MoS_2$ exhibits a simple n-type direct band gap semiconductor, whereas the PL peak shifts towards higher energies and there is no loss in peak intensity relative to pristine $MoS_2$[46]. Although the overall structural phase transition from the 1H to 1T' structure for Re-doped $MoS_2$ appears to take place around 40-50% doping[28,47], our results indicate that the effects on electronic properties induced by dopants happen at far lower concentrations.

**Conclusion and outlook**

We developed sAET to determine the 3D atomic coordinates of 2D materials with picometer precision. We identified 3D crystal defects such as dopants, vacancies and ripples, and measured 3D bond distortion and local strain tensor induced by single dopants. We used the experimental 3D atomic coordinates as direct input to DFT to reveal the electronic band structures of the 2D material, which were corroborated by PL measurements. Compared to other techniques that can image the 3D structure of graphene from only one or two images[36,48], both our experimental and numerical simulation results demonstrate that sAET is a more general method to determine the 3D atomic coordinates of 2D materials and van der Waals heterostructures with high precision. Furthermore, the combination of sAET and ptychography is expected to improve the 3D precision of localizing individual atoms and reduce the electron dose[49,50].

Our study reveals that single dopants in 2D materials can create 3D local bond distortion and induce local strain tensor. Due to the reduced dimensionality, the local strain along the z-axis was measured to be larger than that along the x- and y-axis (Fig. 4). To engineer 2D materials with desired physical, chemical and electronic properties at the atomic scale, it is important to characterize and control the atomic structure in three dimensions. Furthermore, we show the necessity and importance of directly providing experimental 3D atomic coordinates to DFT to reveal electronic properties of a 2D material in metastable states, whereas *ab initio* calculations relying on relaxed 3D atomic models can only predict material properties in the equilibrium state. With the increase of computational power in the future, the complete set of experimental 3D atomic coordinates, determined by sAET and deposited in the Materials Data Bank[51], can be used as direct input to DFT to correlate crystal defects with the electronic, optical, transport and chemical properties of 2D materials and heterostructures. We anticipate that sAET, coupled with DFT and the Materials Data Bank, will not only represent an important advance of how *ab initio* calculations could be used to better understand the structure-property relationship, but also provide feedback to materials engineering at the single-atom level.

**Acknowledgements** We thank M. F. Chisholm for support of the STEM experiment and C. Ophus for help with data analysis. This work was primarily supported by the Office of Basic Energy Sciences of the US DOE (DE-SC0010378). This work was also supported by STROBE: A National Science Foundation Science & Technology Center (DMR-1548924), the Division of Materials Research of the US NSF (DMR-1437263) and an Army Research Office MURI grant on Ab-Initio Solid-State Quantum Materials: Design, Production, and Characterization at the Atomic Scale (18057522). P.M.A. acknowledges support from the Air Force Office of Scientific Research under award number FA9550-18-1-0072. The STEM experiment was conducted by S.Y. with help from J.C.I. at the Center for Nanophase Materials Sciences, which is a DOE Office of Science User Facility.

**Figures and figure legends**

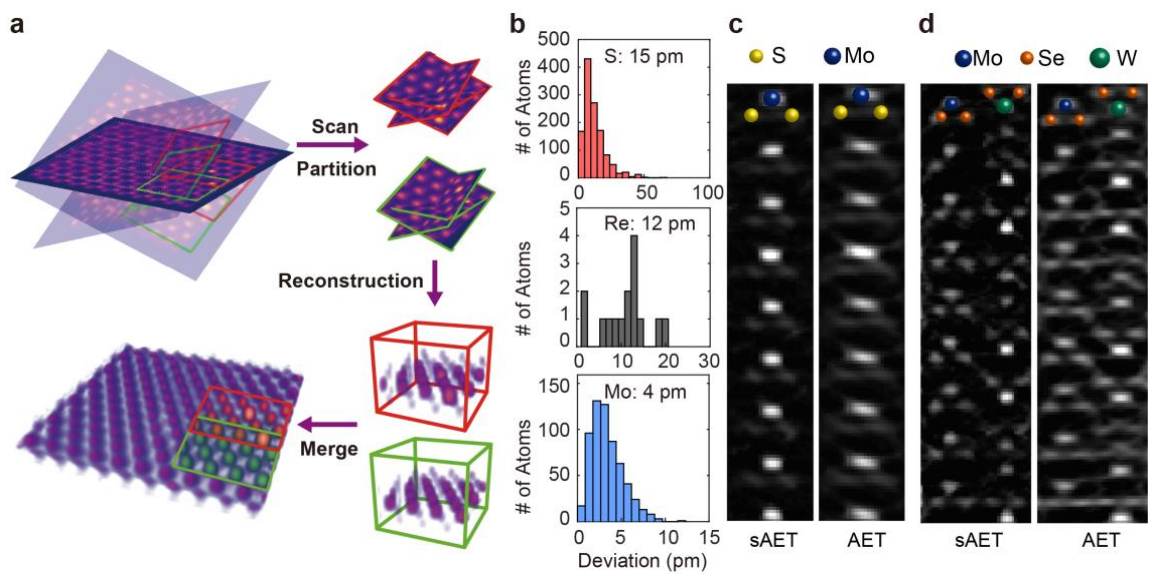

**Figure 1 | Scanning atomic electron tomography (sAET). a,** A limited number of projections are acquired from a 2D material or heterostructure with a low electron dose. A small 3D window is chosen and scanned along the x- and y-axis with overlap between two neighboring windows. At each scanning position, the corresponding regions in the projections are identified for each window. After all the projections are partitioned into a series of image stacks, the 3D windows are reconstructed from the image stacks and stitched together to form a full reconstruction. **b,** The 3D precision of determining the S, Re and Mo atomic positions by sAET was estimated to be 15 pm, 12 pm and 4 pm, respectively, where a total of 1226 S, 15 Re and 611 Mo atoms was used in the statistical analysis. **c,** sAET and AET reconstructions of an experimental Re-doped $MoS_2$ data set, showing that sAET improves the 3D atomic shapes over AET, especially along the z-axis. **d,** sAET and AET reconstructions of a $MoSe_2$-$WSe_2$ heterostructure with moiré patterns from 15 multislice projections. The 3D precision of determining the Mo, W and Se atomic positions by sAET was estimated to be 8 pm, 5 pm and 13 pm, respectively, while the atomic shapes in the AET reconstruction are significantly elongated along the z-axis and are barely traceable.

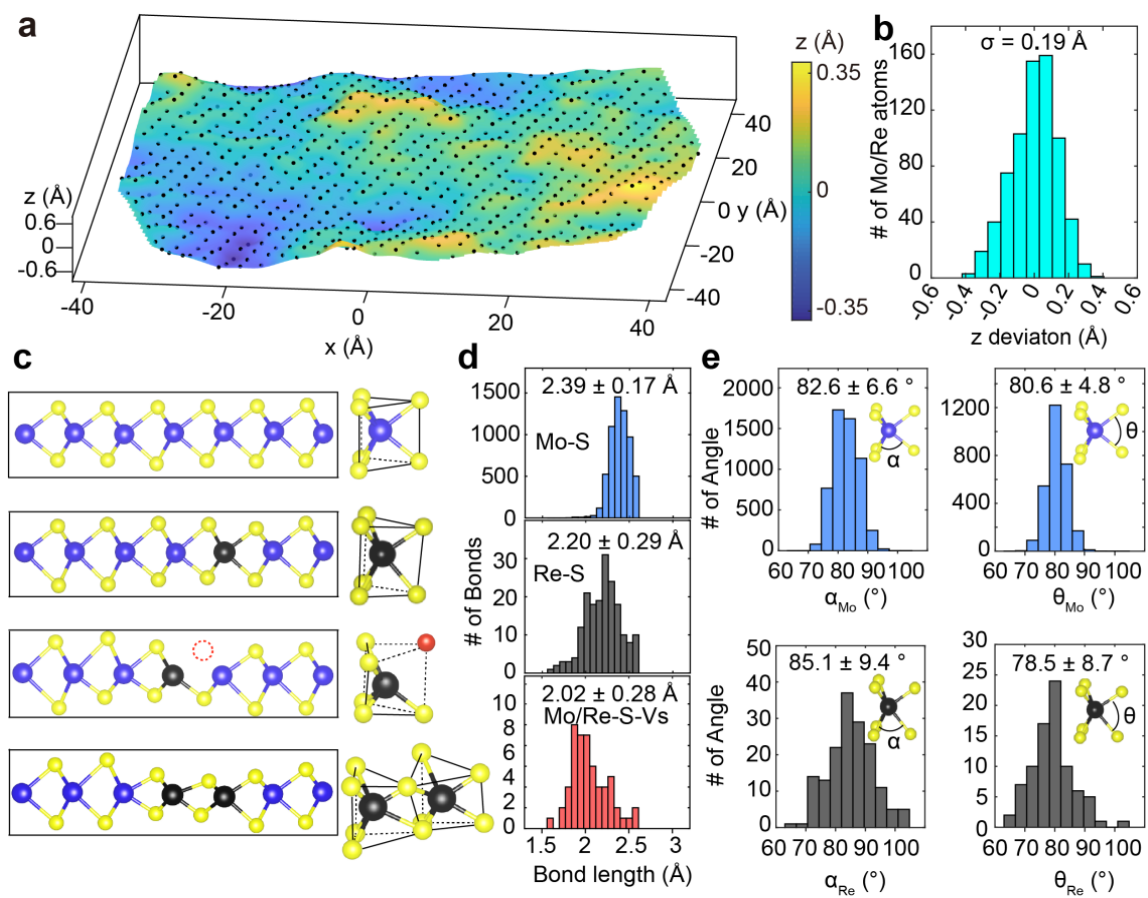

**Figure 2 | 3D atomic coordinates in Re-doped MoS₂ and 3D bond distortion induced by single dopants**. **a**, 3D atomic-scale ripples in data set 1, where the dots represent the Mo/Re atoms. The total number of Re, Mo, S atoms and S vacancies in the data set is 21, 686, 1381 and 15, respectively. **b**, Histogram of the distribution of the z coordinates of the Mo/Re atoms in data set 1 with a standard deviation ($\sigma$) of 19 pm. **c**, Magnified views of four configurations cropped from the 3D coordinates showing the 3D bond distortion: MoS₂ (top), MoS₂ with a single Re dopant (2nd panel), MoS₂ with a Re-vacancy pair (3rd panel) and MoS₂ with double Re dopants (bottom), where the red circle and sphere represent a S vacancy. In each case, the 6 coordinating S atoms (yellow spheres) are shown in the right panels. **d**, Statistical distributions of the Mo-S and Re-S bond length as well as the shortened bond length if there is a S vacancy in the opposite S layer. For reference, the Mo-S bond length of a perfect 1H structure is 2.36 Å. **e**, Statistical distributions the two angles between the M-S bonds (top) and Re-S bonds (bottom) in the same and opposite atomic layer. For reference, the corresponding angles of a perfect 1H structure are $\alpha$ = 83.8°and $\theta$ = 79.1°.

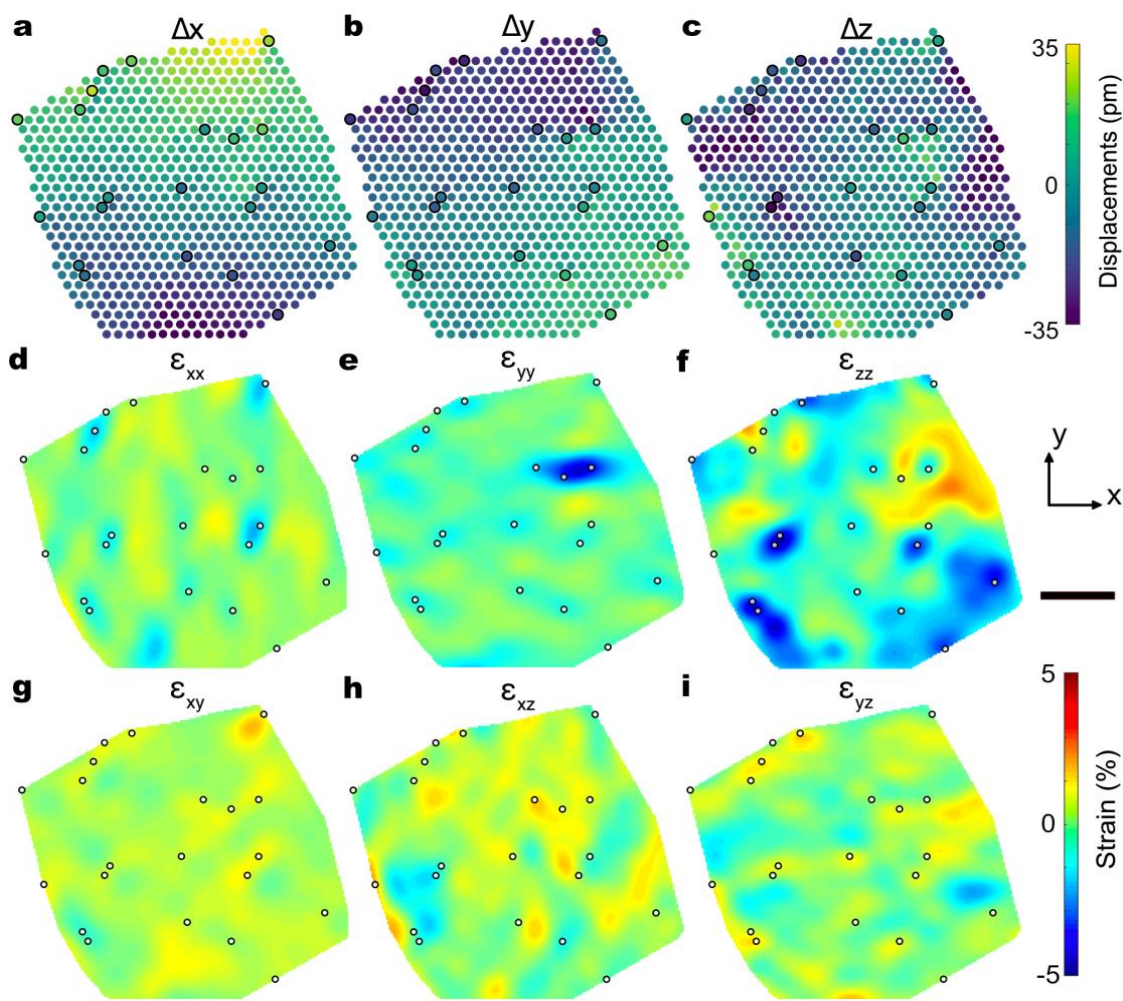

**Figure 3 | Measurements of 3D atomic displacements and the full strain tensor in Re-doped MoS₂. a-c**, 3D atomic displacements of Mo (dots) and Re (circled dots) atoms along the x-, y- and z-axis, respectively. **d-i**, Six components of the strain tensor in the Mo/Re layer, where the Re dopants (circles) induced local strains in the $\varepsilon_{xx}$, $\varepsilon_{yy}$, and $\varepsilon_{zz}$ maps. The x-, y- and z-axis are along the [1,0,0], [1,2,0] and [0,0,1] directions, respectively. The strain tensor of the two S layers is shown in Supplementary Fig. 10. Scale bar, 2 nm.

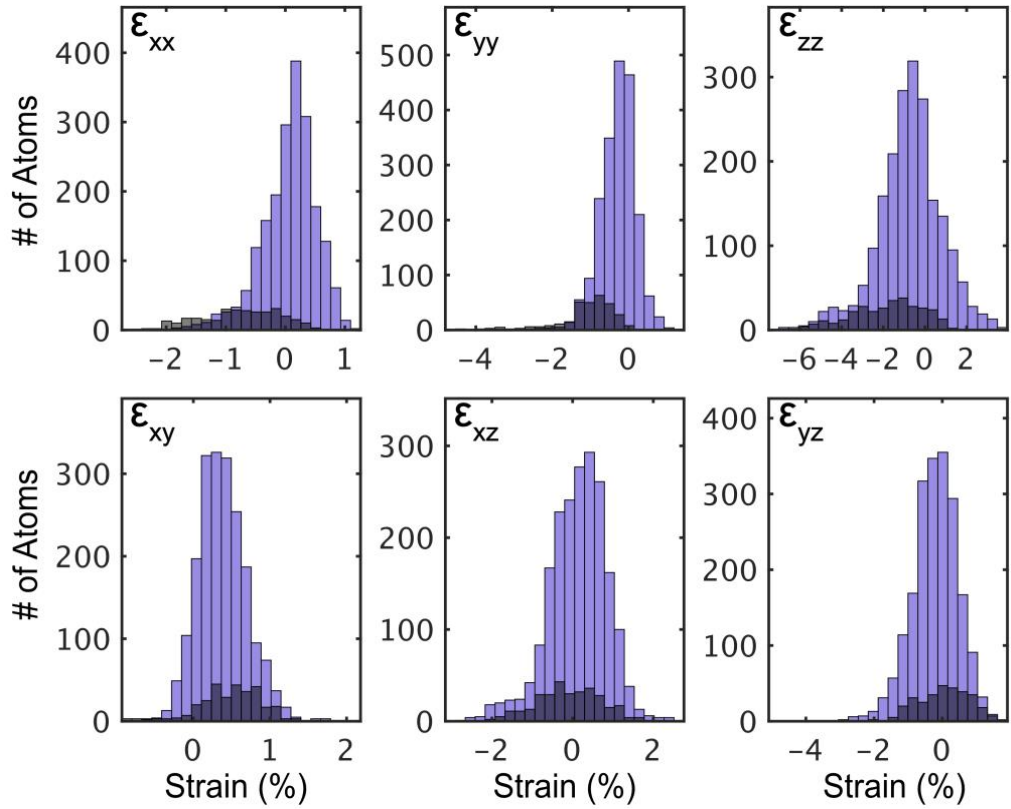

**Figure 4 | Measurements of the local strain tenor induced by single Re dopants.** Spheres with a radius of 3.16 Å were chosen centered at the Re and Mo atoms. On average the histogram distributions (purple) of the local strain tensor at atomic positions within the Mo spheres were determined to be $\varepsilon_{xx}$ = 0.06±0.45, $\varepsilon_{yy}$ = -0.28±0.48, $\varepsilon_{zz}$ = -0.73±1.54, $\varepsilon_{xy}$ = 0.38±0.32, $\varepsilon_{xz}$ = 0.12±0.72, and $\varepsilon_{yz}$ = -0.21±0.69, whereas the histogram distributions (black) of the local strain tensor at atomic positions within the Re spheres were estimated to be $\varepsilon_{xx}$ = -0.76±0.67, $\varepsilon_{yy}$ = -1.1±0.77, $\varepsilon_{zz}$ = -1.88±1.75, $\varepsilon_{xy}$ = 0.41±0.49, $\varepsilon_{xz}$ = -0.1±0.78, and $\varepsilon_{yz}$ = 0.02±0.77.

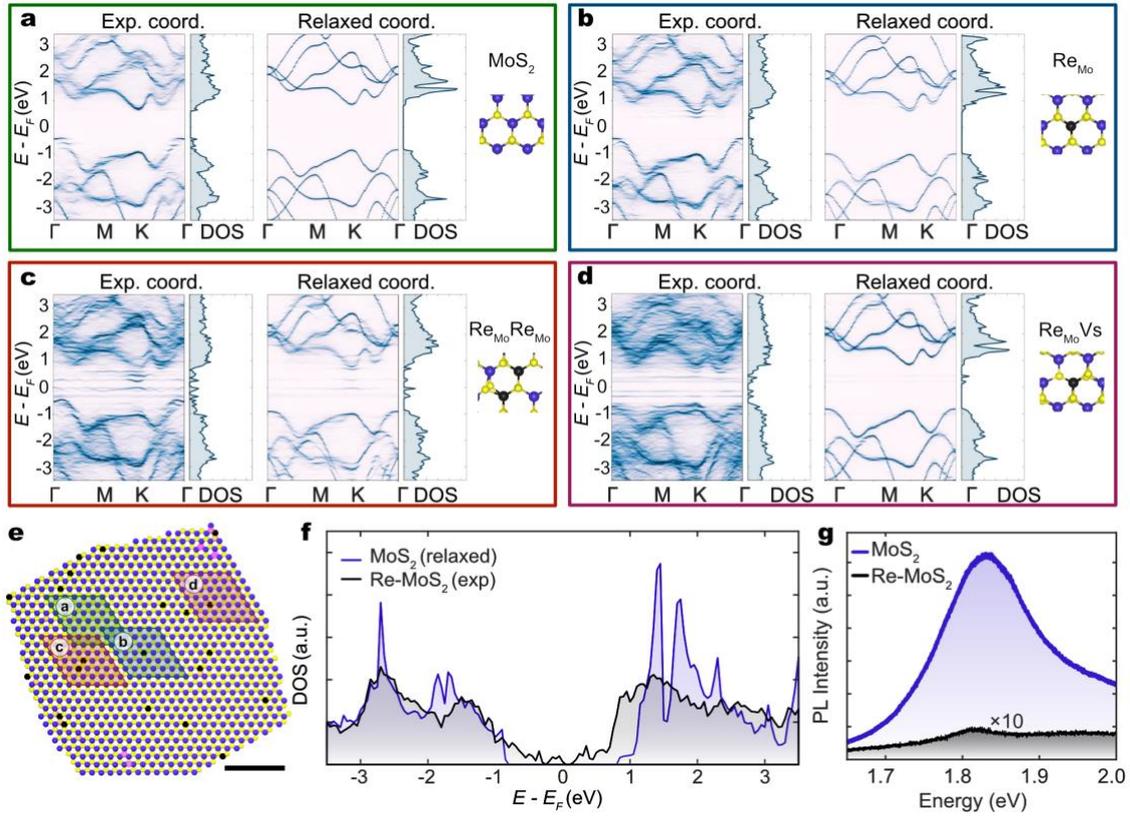

**Figure 5 | Electronic band structures derived from experimental 3D atomic coordinates and photoluminescence measurements.** Four supercells selected from different regions of Re-doped $MoS_2$ (**e**), including a dopant-free $MoS_2$ structure (**a**), a single Re dopant (**b**), double Re dopants (**c**), and a Re dopant and S vacancy (**d**). The experimental coordinates of the supercells were used as direct input to DFT to reveal the electronic band structures (**a-d**, left). For comparison, the same experimental coordinates were relaxed to obtain the band structures by DFT (**a-d**, middle). **f,** Average DOS of the experimental structures in the four supercells (black curve) and relaxed $MoS_2$ DOS (blue curve). **g,** Photoluminescence (PL) spectra of as synthesized pristine $MoS_2$ (blue) and Re-doped $MoS_2$ monolayers (black). Scale bar, 2 nm.

## METHODS

**2D materials synthesis.** Molybdenum oxide powder (99%, Sigma Aldrich), sulfur powder (99.5%, Sigma Aldrich) and ammonium perrhenate (99%, Sigma Aldrich) were used as precursors for chemical vapor deposition growth[52]. A selected ratio of molybdenum oxide and ammonium perrhenate was added to an alumina boat with a Si/SiO₂ (285 nm) wafer cover. The furnace temperature was ramped to 550 °C in 15 min and then kept at 550 °C for another 15 min for the growth of the Re-doped $MoS_2$ alloy materials. We found the films to show a doping concentration of roughly 3%. Sulfur powder in another alumina boat was placed upstream where the temperature was roughly 200 °C. After growth, the furnace was cooled to room temperature using natural convection. The growth process was carried out under 50 SCCM argon at atmospheric pressure. The Re-doped $MoS_2$ flakes were transferred to 3 mm Quantifoil TEM grids by spin coating the samples with PMMA to support the flakes, and then were etched with KOH to release the flakes from the substrates. The sample was dipped into Acetone for 2 hours to wash away PMMA and then baked in vacuum at $10^{-3}$ Pa and 160 °C.

**Low-dose data acquisition of 2D materials.** Two data sets, each consisting of 13 projections, were acquired from different regions of a Re-doped $MoS_2$ monolayer using a Nion UltraSTEM 100[TM] aberration-corrected scanning transmission electron microscope (STEM) (Supplementary Table 1). Images were collected at 60 kV in annular dark-field (ADF)-STEM mode[52,53]. The beam convergence semi-angle was 30 mrad and the detector collection angle was in the range of 30-300 mrad, where a small detector inner angle was chosen to reduce the electron dose. The energy spread of the electron beam was 0.3 eV. A double tilt sample holder was used to collect the data sets (Supplementary Table 2). To reduce the total electron dose, 10 images per angle were measured with a beam current of 15 pA and a dwell time of 4 μs per image. To monitor any potential structural changes induced by the electron beam, the first and last projections of each data set were compared to ensure that no noticeable structural changes were observed during data acquisition (Supplementary Fig. 2). The total electron dose of each data set was estimated to be $4.1 \times 10^5$ e⁻/Å² (Supplementary Table 1).

**Drift correction.** To reduce the sample drift effect on data acquisition, we took 10 STEM images for each tilt angle and aligned them using the following procedure. First, we chose a region of $300 \times 300$ pixels from the first image as a reference. Next, a region of $200 \times 200$ pixels from the next image was cropped and scanned over the reference image with a step size of 0.1 pixel to calculate the cross-correlation coefficients. The maximum cross correlation coefficient corresponds to the relative drift between the two images. Typically, the drift between two neighbouring images is less than 1 pixel. By repeating this process, the relative drift for all ten images was determined and an overall drift vector was calculated. With the drift vector, we applied scan distortion correction by assuming the drift vector is uniformly distributed along the slow scan direction during STEM image acquisition. Thus, the drift for each pixel in the image was determined and a drift corrected image was formed by interpolating the non-corrected image onto the drift corrected pixel positions. After applying drift correction, we averaged the 10 images to obtain a final image for each tilt angle.

**Image denoising**. The raw ADF-STEM images contain mixed Poisson and Gaussian noise, and a Block-matching and 3D filtering (BM3D) algorithm was applied to denoise the average image of each tilt angle[54]. Two different sets of denoising parameters were applied for the BM3D: one is the exact noise level estimated from the experimental images (termed BM3D 100%), and the other is by doubling the estimated noise level (termed BM3D 200%). There are two reasons for the use of BM3D 100% and 200%. First, this would allow us to cross-validate the reconstructions with different denoising parameters. Second, the BM3D 100% reconstruction provided better contrast between the Re, Mo and S atoms, while BM3D 200% produced better reconstruction for the S atoms. All these post-processing and denoising methods have been previously demonstrated to be robust for dealing with ADF-STEM images[17, 19-21].

**Angle calibration and 3D coordinate fitting of the Re and Mo atoms.** The nominal angles were measured by the Nion double tilt stage (Supplementary Table 2). After denoising each projection, we determined the x and y coordinates of each Re and Mo atom by fitting a 2D Gaussian in a 5×5 pixel region. According to a geometric relation, the x and y coordinates of the Re and Mo atoms in each data set change as a function of the tilt angle. To calibrate the tilt angles and determine the 3D coordinates of the Re and Mo atoms, we used the least square method to minimize the following equation,

$$E = \sum_i \sum_j \left\{ \left[ P_x(\boldsymbol{r}_i, \alpha_j, \beta_j) - x_i^j \right]^2 + \left[ P_y(\boldsymbol{r}_i, \alpha_j, \beta_j) - y_i^j \right]^2 \right\} \qquad (1)$$

where $P_x$ and $P_y$ are the functions of projecting the 3D coordinates of the Re and Mo atoms to the x and y coordinates in the projections, respectively, $\boldsymbol{r}_i$ is the 3D coordinates of the $i^{th}$ Re or Mo atom, $\alpha_j$ and $\beta_j$ are the tilt angles of the $j^{th}$ projection, $x_i^j$ and $y_i^j$ are the measured x and y coordinates in the $i^{th}$ atom in the $j^{th}$ projection, respectively. By minimizing $E$, we calibrated the tilt angles of the 13 projections (Supplementary Table 2) and obtained the 3D coordinates of all the Re and Mo atoms in each data set (Supplementary Fig. 7). This least square fitting method is robust and produces consistent results regardless of different initial input.

**Deconvolution for vibrational correction.** Due to the high length/thickness ratio and the free-standing structure, we observed the 2D material suffered from vibrational blurring along the perpendicular direction to the TEM grid during data acquisition. The vibrational blurring is equivalent to convolving the images with a kernel and can be removed through deconvolution if the blurring kernel is known. In this experiment, the blurring kernel was estimated based on the fact that all atoms should be spherical. We performed deconvolution using the following procedure. First, we interpolated each experimental image by a factor of 2 with linear interpolation and cropped a region of 100×100 pixels without Re dopants. Second, a $MoS_2$ model of the same experimental tilt angle was used to create a reference image of the same size as the cropped experimental image but with spherical atoms. The reference image was normalized and aligned with the experimental image. Third, a vibrational kernel was constructed by adjusting the vibrational direction, vibration amplitude and Gaussian blurring. The experimental image was deconvolved with the constructed kernel using the Lucy-Richardson algorithm[55,56] and compared with the reference image. A brute-force process was conducted until an optimal kernel was obtained, creating the best match between the experimental and reference images. Finally, the optimal kernel was applied to the whole experimental image using the Lucy-Richardson algorithm. The deconvolved image was binned to be its original scale. A comparison between before and after deconvolution is shown in Supplementary Fig. 3.

**Image partition and reconstruction with sAET.** To implement sAET, we chose a 3D window of 60×60×60 voxels and scanned it along the x and y axis with a step size of 30 voxels. This step size offsets each partition to include a 30 voxel overlap between neighboring partitions on each side. At each step, the corresponding regions from all 13 projections in each data set were cropped and grouped into an image stack. Each image stack consists of 13 images with varied shapes, corresponding to the projection of the 3D window along different tilt angles. After a full 2D scan was completed, all the projections were partitioned into image stacks. All the image stacks were aligned and reconstructed in parallel by the generalized Fourier iterative reconstruction (GENFIRE) algorithm[30]. Each GENFIRE reconstruction used a 33-voxel support along the z-axis and ran 1000 iterations. Due to the extended nature of the 2D materials along the x and y axis, the reconstruction of each image stack contained artifacts near the boundary. To remove these artifacts, we stitched together only the central 30×30×33 voxels of the reconstructed windows to produce a full 3D reconstruction. By applying this partition and reconstruction procedure, we obtained the full 3D reconstructions of both the BM3D 100% and 200% projections of each data set.

**Initial localization of 3D atomic coordinates and species.** The 3D atomic coordinates and species of the 2D material were initially traced from the 3D reconstructions using the following procedure. We first identified all local maxima in the BM3D 200% reconstruction of each data set. Starting from the highest-intensity local maximum peak, we cropped a $1.71×1.71×1.71Å^3$ (5×5×5 voxel) volume with the selected local peak as the center. We fitted the volume with a 3D Gaussian function described elsewhere[19-21]. If a fitted peak position satisfied a minimum distance of 1.6 Å away from any previously fitted peak (i.e. a minimum distance constraint), we added it to a list of potential atoms. By applying the 3D Gaussian fitting algorithm to all the identified atoms, we obtained a complete list of potential atom positions. These positions were manually checked to correct for unidentified or misidentified atoms due to fitting failure or areas with connected intensity blobs from multiple atoms. We then assigned the atomic species based on the 3D intensity distribution of the traced potential Re, Mo and S atoms.

**Refinement of 3D atomic coordinates and species and identification of S vacancies.** The traced 3D atomic coordinates and species were refined by the following procedure. First, each experimental image of BM3D 100% was converted to Fourier slices $F_{obs}^j(\boldsymbol{q})$ with $j = 1,\ldots,13$, by the fast Fourier transform. Next, 13 Fourier slices were calculated with the atomic model by

$$F_{calc}^j(\boldsymbol{q}) = \sum_{n=1}^{N} H f_e(q) e^{\frac{-B'q^2}{4} - 2\pi i r_n \cdot \boldsymbol{q}}, \qquad (2)$$

where $N$ is the number of atoms, $H$ is the scaling factor for different atomic species, the $B'$ factor accounts for the electron probe size, the missing wedge and the thermal motion of each atomic species, $f_e(q)$ is a normalized electron scattering factor, $\boldsymbol{r}_n$ is the 3D position of the $n^{\text{th}}$ atom. An error function was then calculated by

$$E = \sum_{j=1}^{13} \sum_{q} \left| F_{obs}^j(\boldsymbol{q}) - F_{calc}^j(\boldsymbol{q}) \right|^2, \qquad (3)$$

which was minimized with respect to the atomic positions ($\boldsymbol{r}_n$) by a gradient descent method[19-21].

From the refined 3D atomic model, we developed an atom pair flipping method to identify the S vacancies, which consists of the following four steps. First, we randomly chose a pair of S atoms between the top and bottom atomic layers. For each selected S pair, projection images were calculated for all the tilt angles by flipping the pair among four cases: i) both atoms are S, ii) both are vacancies, iii) the top is a S atom and the bottom is a vacancy, and iv) the top is a vacancy and the bottom is a S atom. Four atomic models were generated accordingly to the four cases, with the same $H$ and $B'$ factors. Second, four sets of 13 projections with the experimental tilt angles were calculated from the atomic models. An $R_1$ factor was computed between measured and calculated projections,

$$R_1^j = \frac{\sum_{x,y} \left| f_{obs}^j(x,y) - f_{calc}^j(x,y) \right|}{\sum_{x,y} \left| f_{obs}^j(x,y) \right|} , \qquad (4)$$

$$R_1 = \frac{\sum_j R_1^j}{13} , \qquad (5)$$

where $f_{obs}^j(x,y)$ and $f_{calc}^j(x,y)$ are the $j^{th}$ measured and calculated projection. As flipping a pair of atoms only affects a small cylindrical volume for each tilt angle, we only calculated a small area of projection along a cylindrical volume containing the atom pair. By comparing the $R_1$ factor among the four cases, the one with the smallest error $R_1$ was chosen and updated in the atomic model. Third, we repeated the first two steps for all the S atom pairs and obtained an updated 3D atomic model. A global scale factor was calculated for the updated atomic model to minimize the error between the measured and calculated projections. Fourth, we iterated steps one to three for all the S atom pairs until there was no change in the atomic species. The atom pair flipping method was robust and converged after a few iterations for both data sets. After identifying all the S vacancies, we refined the 3D atomic coordinates once more using equations (2) and (3).

**Dynamic refinement of the S atoms near the Re atoms.** Due to the use of a low energy electron beam (60 keV) and a small detector inner angle to reduce the electron dose, we observed the dynamic scattering effect of the Re atoms (see the next section in detail), which influenced some nearby S atoms. We implemented a brute-force approach to perform dynamic refinement of these Re and S atoms. We first identified those S atoms nearby the Re atoms, which either violated a minimum distance of 1.6 Å (caused by the refinement) or deviated from the S atomic layer more than 1 Å. There were 11 and 13 such S atoms in data set 1 and 2, respectively. For each of these S and Re atoms, we performed a 3D scan of its position with a range of ±40 pm along the x and y axis and ±120 pm along the z axis. The step size is 20 pm, 20 pm and 30 pm along the x, y and z axis, respectively. At each scanning step, we used multislice simulations to calculate 13 projection images of 40×40 pixels in size at different tilt angles and computed the $R_I$ factor relative to the measured images. The S atom in the atomic model was updated to the position corresponding to the minimum $R_I$ factor. We repeated this dynamic refinement procedure for all the S atoms whose positions were influenced by nearby Re atoms.

**Multislice simulations for precision estimation.** We performed multislice simulations to estimate the 3D precision of the atomic coordinates and species[57,58]. The experimental atomic model was placed in a rectangular super cell. The super cell was divided into multiple 2 Å slices along the z axis, and the x–y plane was discretized into 1,920×1,920 pixels. The same experimental parameters were used for the multislice simulations (E: 60 keV; C3: 0 mm; C5: 5 mm; convergence semi-angle: 30 mrad; detector inner angle: 30 mrad and detector outer angle: 300 mrad). Each STEM image was generated by a raster scan in the x-y plane with a step size of 0.34 Å. For each tilt angle, we averaged 12 phonon configurations to obtain a projection image. The image was convolved with a 5×5 pixel Gaussian kernel ($\sigma = 1.0$) to account for the electron probe size, thermal vibrations, and other incoherent effects. The mixture of Gaussian and Poisson noise determined from the experimental images was added to the multislice images (Supplementary Fig. 8). From the 13 multislice images of data set 1, we used the same imaging processing, sAET reconstruction, atom tracing and refinement procedures to obtain a new 3D atomic model. By comparing it with the experimental 3D atomic model, we found all the atom species were correctly identified including all S vacancies. The RMSD of the S, Re and Mo atoms is 15 pm, 12 pm and 4 pm, respectively (Fig. 1b).

Next, we performed multislice simulations using the same parameters except changing the detector inner angle to 50 mrad. We added Gaussian and Poisson noise to the multislice projections with the same electron dose and obtained another 3D atomic model. In comparison to the experimental atomic model, all the atom species were correctly identified and the RMSD of the S, Re and Mo atoms is 18 pm, 3 pm and 3 pm, respectively. Compared to the 30 mrad detector inner angle results, the RMSD of the Re atoms is decreased due to the reduced dynamic scattering effect, while the RMSD of the S atoms is increased due to the weaker intensity in the 50 mrad detector inner angle case.

**Measurement of the 3D strain tensor of Re-doped MoS₂.** The strain tensor was determined from the experimental 3D atomic coordinates using a procedure described elsewhere[19]. The experimental atomic model of each data set was least-square fitted to an ideal $MoS_2$ model. The displacement vectors, $\Delta \boldsymbol{u}$, were calculated as the difference in the atomic positions between the experimental and ideal model. Each displacement component ($\Delta x$, $\Delta y$, $\Delta z$) was interpolated to a cubic grid through kernel density estimation[59]. A 3D Gaussian kernel with $\sigma = 3.16$ Å was convolved with the displacement fields, which increases the precision of the strain tensor measurement, but reduces the resolution. The six components of the strain tensor of the two data sets are shown in Supplementary Fig. 10.

To determine the principal strain of the four supercells in Fig. 5, we calculated the average strain tensor within each supercell and then solved for the eigenvalues and eigenvectors of the average strain tensor. The principal strain and three corresponding directions are [0.4%, -0.23%, -1.15%], [0.86,0.20,0.47], [0.15,0.79,-0.60] and [-0.49,0.58,0.65] for supercell A, [0.5% -1.00%, -0.60%], [-0.87,-0.37,-0.32], [-0.48,0.75,0.44] and [0.08,0.54,-0.84] for supercell B, [0.14%, -0.71%, -1.58%], [0.93,0.37,0.04], [-0.37,0.92,0.10] and [0.00,-0.11,0.99] for supercell C, and [-0.04%, -1.84%, 1.35%], [-0.92,-0.20,0.34], [-0.26,0.95,-0.15] and [0.30,0.23,0.92] for supercell D.

**DFT calculations of the effective band structure.** To perform DFT calculations of the electronic properties of the Re-doped $MoS_2$ materials from the experimental 3D coordinates, a 20 Hartree plane-wave energy cutoff was applied on a 3×3×1 K-point mesh on 6×6×1 supercells of the $MoS_2$ sub-regions. We employed ultrasoft pseudopotentials and modeled the exchange and correlation interactions with the PBEsol functional. A Fermi smearing of 0.01 Hartrees was used to more accurately account for temperature effects present during experimental measurements at room temperature. In order to adequately describe the isolated monolayer without neighboring image effects, a Coulomb truncation scheme was used in the out-of-plane direction[60]. All calculations were performed using JDFTx[61].

**Band Unfolding.** Due to the relationship between reciprocal and real space, the Brillouin zone of a supercell is some fraction of the size of a primitive cell Brillouin zone. To recover information about the cell with respect to a reference primitive cell, one must apply band unfolding techniques to the Kohn-Sham wavefunctions[62,63], which were used to obtain the effective band structures shown in Fig. 5. The reciprocal spaces of a primitive unit cell and a supercell are geometrically related, with the lattice vectors of one being an integer matrix transformation of the other. We unfolded each cell onto a primitive cell with a hexagonal lattice constant equal to the experimental value. Because we modeled each subsystem as a supercell, periodic boundary conditions were applied to a larger real space lattice than in a traditional primitive cell. As a result, the Brillouin zone of our sub-regions is smaller than that of a primitive cell hexagonal lattice, and the high symmetry k point path differs between the two.

This leads naturally to a description of the supercell Bloch states in terms of the primitive cell $\boldsymbol{k}$ states for which $\boldsymbol{k}$ in the primitive cell is equivalent to $\boldsymbol{K}$ in the super cell by a reciprocal lattice vector translation. In particular, the overlap of the supercell Bloch states (indexed by a supercell wave vector $\boldsymbol{K}$ and band index $m$) with the Bloch states of a primitive cell is given by the spectral weight:

$$P_{\vec{K}m}(\boldsymbol{k}_i) = \sum_n |\langle \boldsymbol{K}m|\boldsymbol{k}_i n\rangle|^2 \quad . \qquad (6)$$

where the supercell $\boldsymbol{K}$-point is projected (unfolded) to the $\boldsymbol{k}_i$-points in the primitive cell. In practice, the spectral weights can be obtained without reference to the primitive cell Bloch states, provided the matrix is known which transforms the desired primitive cell lattice to that of the supercell. The spectral weights are used to find the spectral function, defined by:

$$A(\boldsymbol{k}_i, E) = \sum_n P_{\boldsymbol{K}m}(\boldsymbol{k}_i)\delta(E_m - E) \ . \qquad (7)$$

In the effective band structure, the relative weight of the spectral function is referenced via the color map intensity for each point in the primitive cell Brillouin zone. The formalism for this is principally exact. The spectral function technique directly quantifies the degree to which the primitive cell symmetry is maintained in the supercell. If a state with wave vector $\boldsymbol{K}$ at some energy does not possess any of the symmetries of the primitive cell, the spectral weight of these points will be zero, such that they do not appear in the resulting band structure. For the case of a supercell constructed purely from some reference primitive cell, the unfolding procedure will produce the expected band structure exactly because the supercell possesses all of the same higher symmetries as the primitive cell, and the spectral weights will all be unity.

On the other hand, disordered, strained, or otherwise non-perfect supercells will possess shadow bands or broadening: points of the primitive cell Brillouin zone where the spectral weights have intermediate values between 0 and 1. This indicates that some, but not all, of the underlying symmetries are preserved. While these methods should not be expected to exactly reproduce standard results in these situations, these features can be useful in understanding the nature of a given cell.

**Photoluminescence and Raman spectra measurements.** The photoluminescence spectroscopy experiment was performed at room temperature using a Renishaw inVia instrument. Both pristine $MoS_2$ and Re-doped $MoS_2$ monolayer samples were synthesized under the same conditions and supported on $SiO_2$ substrates. An excitation laser of 514.5 nm wavelengths was focused onto the samples via a 100X objective lens, producing a spot diameter of ~1 µm. The laser power was kept at 5 mW. Background measurements on the substrate are shown in Supplementary Fig. 12a. In many photoluminescence measurements we found the Re-doped $MoS_2$ spectra were quenched, with no visible signal above the background. Supplementary Fig. 12b shows the raw spectrum from a Re-doped $MoS_2$ region with a small exciton peak (black), the background measurement (red), and the background subtracted spectrum (shown in Fig. 5g). The large difference in scale between the pristine and undoped $MoS_2$ photoluminescence spectra is clearly seen in Supplementary Fig. 12c. The photoluminescence spectra show the $MoS_2$ monolayer is a direct band gap semiconductor and corroborate that the Re-doped $MoS_2$ monolayer is a highly distorted indirect band gap semiconductor or metal-like behaviour (Fig. 5f).

We also measured the Raman spectra from the pristine $MoS_2$ and Re-doped $MoS_2$ monolayer samples supported on $SiO_2$ substrates. Using the peak position and peak intensity of Si at 521 cm$^{-1}$ as a reference, we observed that, for both the in-plane ($E_{2g}$) and out-of-plane ($A_{1g}$) phonon modes, the peak intensity from Re-doped $MoS_2$ is about 1/3 of that from pristine $MoS_2$ (Supplementary Fig. 13). The peak intensity reduction in Re-doped $MoS_2$ suggests the lattice of the 1H $MoS_2$ structure was significantly distorted by Re dopants, which was confirmed by our direct experimental observations in three dimensions (Figs. 2-4). With the further increase of Re dopants (> 4.4%), both the $E_{2g}$ and $A_{1g}$ mode peaks disappeared in the Raman spectrum[64], indicating the loss of the long-range order in 1H structure of $MoS_2$.

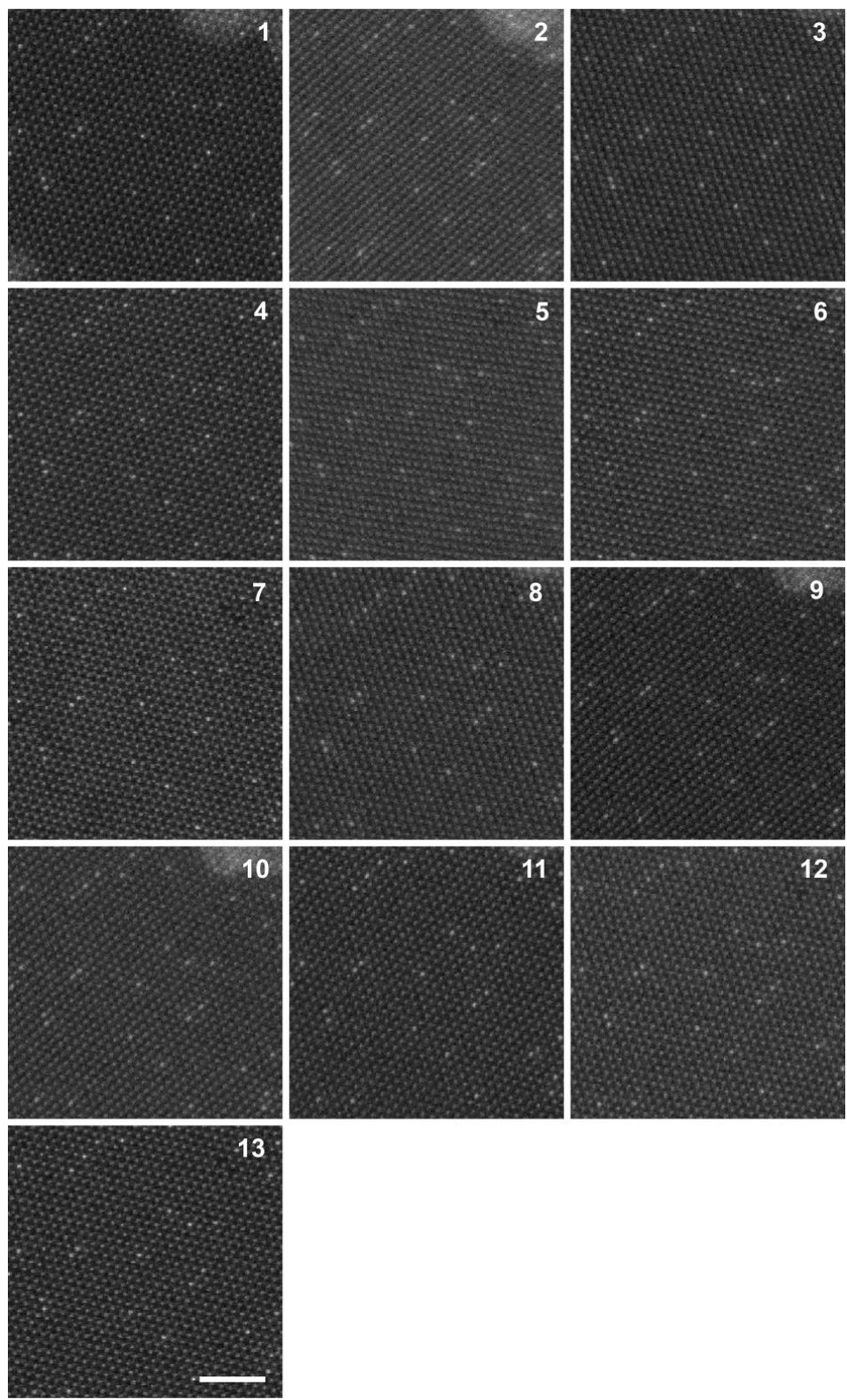

**Supplementary Figure 1 |** A tilt series of 13 experimental projections of a Re-doped $MoS_2$ monolayer (data set 1). The bright dots are individual Re dopants. Scale bar, 2 nm.

First image in tilt series　　　　　Last image in tilt series

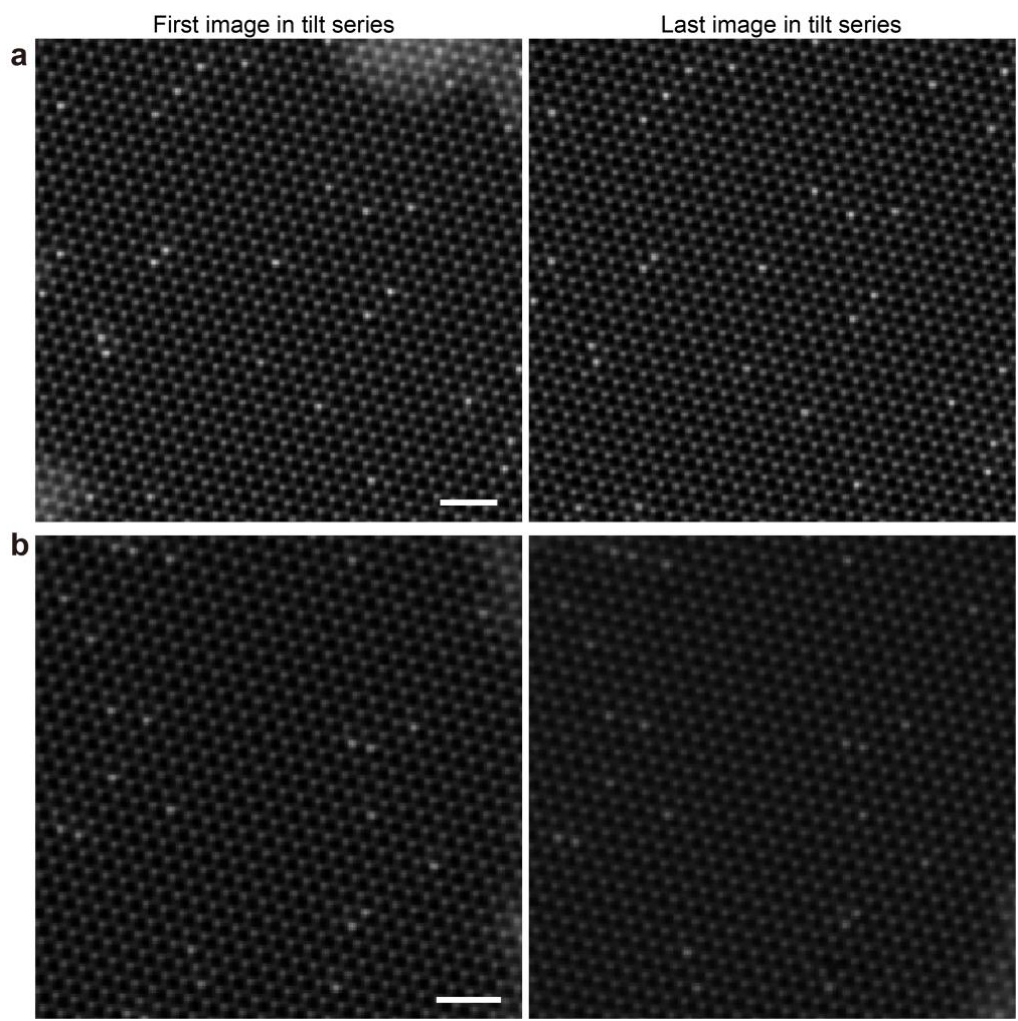

**Supplementary Figure 2 | Sample Consistency check. a** and **b**, Comparison of the first and last images for data set 1 and data set 2, respectively, indicating the consistency of the atomic structure during data acquisition. Scale bar, 1 nm.

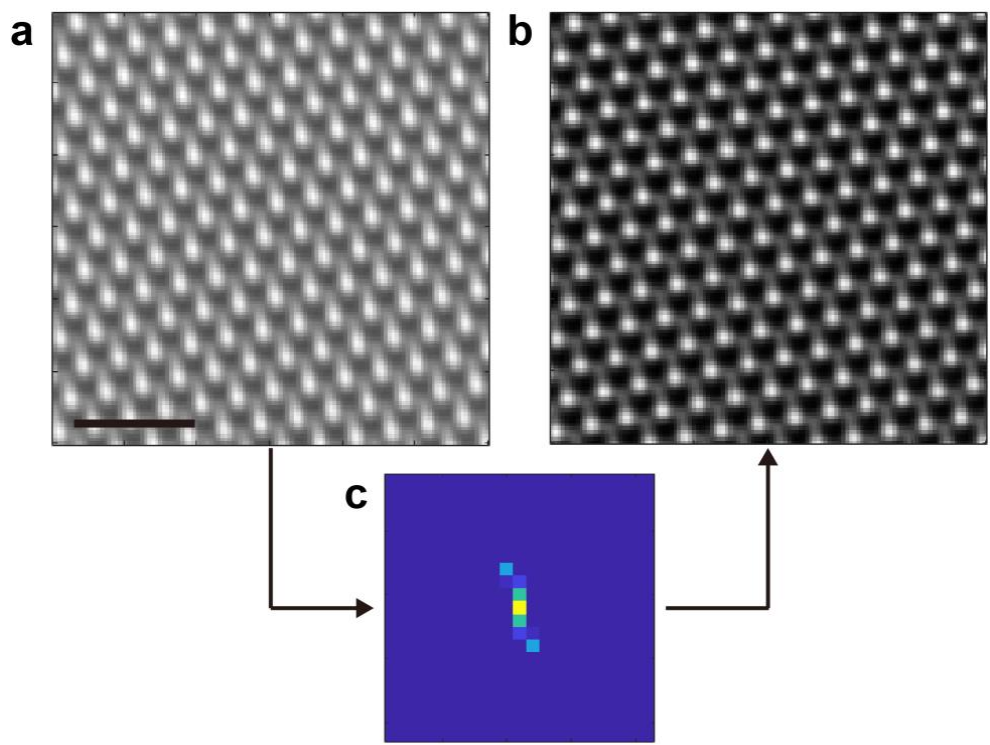

**Supplementary Figure 3 | Image deconvolution for sample vibration correction. a,** A representative image of the Re-doped $MoS_2$ monolayer at a high tilt angle, where the atoms are elongated due to vibrational blurring. **b,** The same image after deconvolution using a harmonic oscillator kernel, where the atoms become spherical. **c,** The harmonic oscillator deconvolution kernel. Scale bar, 1 nm.

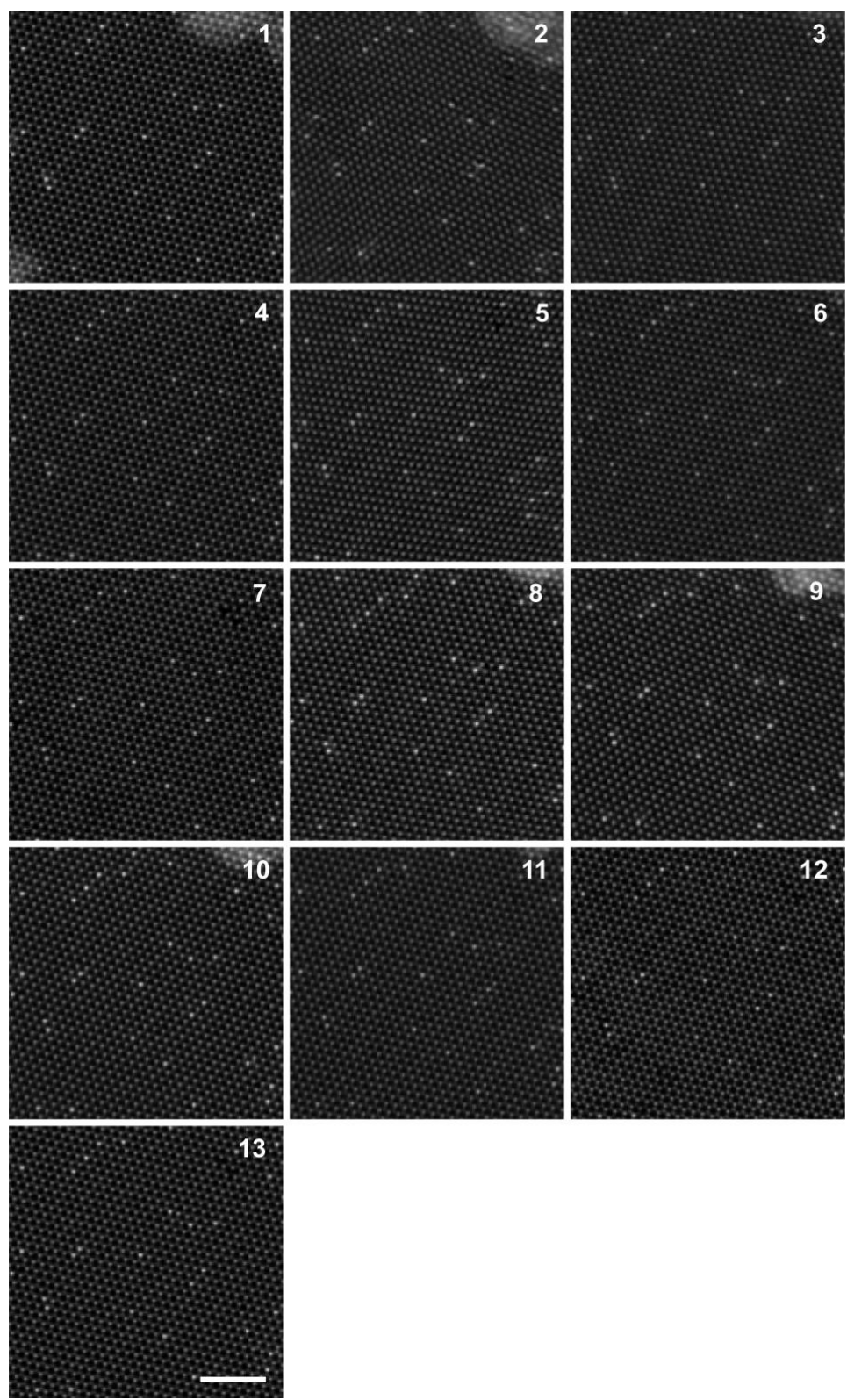

**Supplementary Figure 4 |** The same tilt series of 13 experimental projections shown in Fig. 1 after denoising and deconvolution. Scale bar, 2 nm.

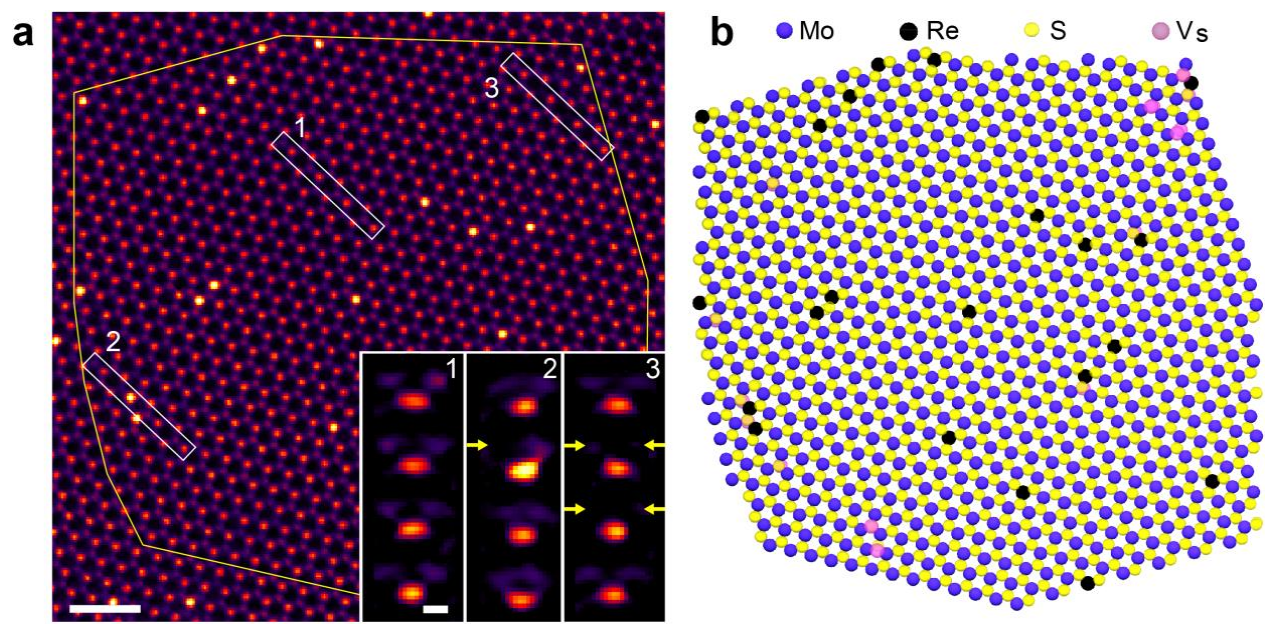

**Supplementary Figure 5 | 3D atomic coordinates and crystal defects in data set 1. a,** Top view of the 3D reconstruction of data set 1. The inset shows the side view of MoS$_2$ (panel 1), MoS$_2$ with a Re dopant and a S vacancy (panel 2), and MoS$_2$ with 3 S vacancies (panel 3), where arrows indicate the vacancies. Scale bar, 1 nm and scale bar (inset), 2 Å. **b,** 3D atomic model of the bounded region in (**a**), consisting of 1381 S, 686 Mo and 21 Re atoms with 15 S vacancies.

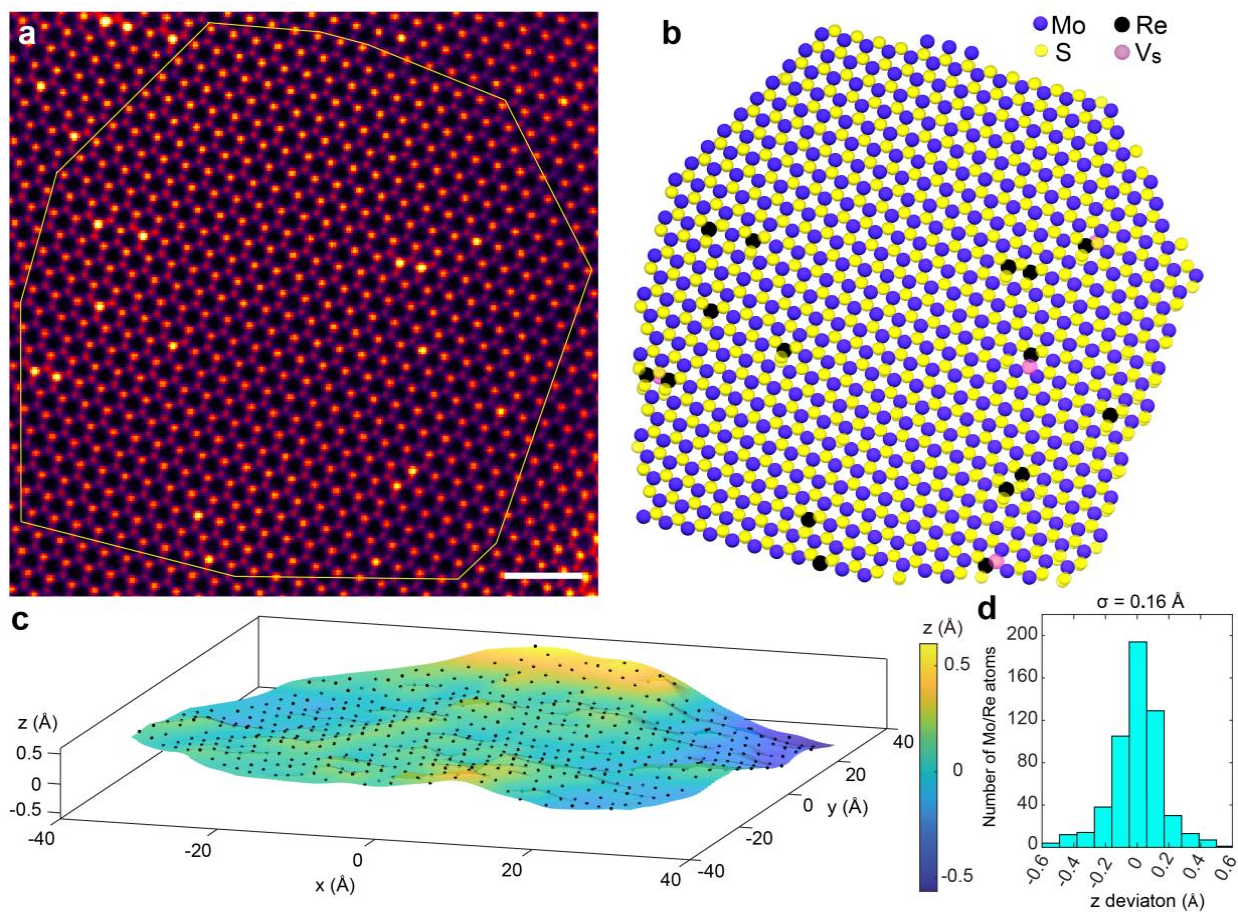

**Supplementary Figure 6 | 3D atomic coordinates and crystal defects in data set 2. a,** Top view of the 3D reconstruction. Scale bar, 1 nm. **b,** 3D atomic model of the bounded region in (**a**), consisting of 1083 S, 531 Mo and 16 Re atoms with 4 S vacancies. **c,** 3D plot of the Mo/Re layer showing atomics-scale ripples, where the dots represent the Mo/Re atoms. **d,** Histogram of the distribution of the z coordinates of the Mo/Re atoms in data set 1 with a standard deviation ($\sigma$) of 0.16 Å.

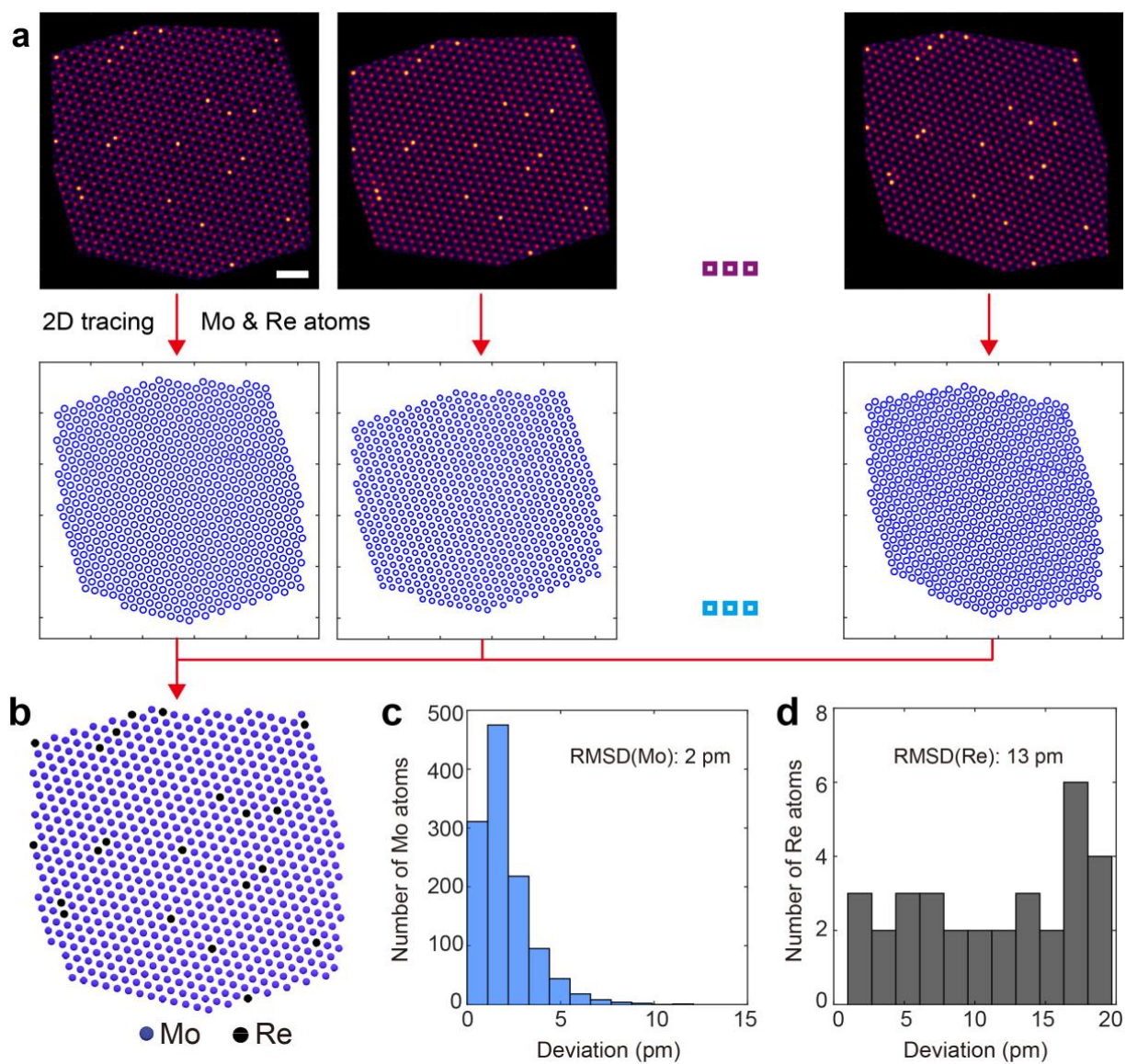

**Supplementary Figure 7 | Least square fitting to determine the 3D coordinates of the Re and Mo atoms. a,** The x and y coordinates of the Re and Mo atoms were localized from the aligned projections of each data set. **b,** From these 2D coordinates, the tilt angles were calibrated and the 3D coordinates of the Re and Mo atoms were determined for each data set. The 3D atomic coordinates are consistent with those obtained by sAET with a RMSD of 2 pm and 13 pm for Mo (**c**) and Re (**d**) atoms, respectively, where a total of 1176 Mo and 32 Re atoms was used in the statistical analysis. Scale bar, 1 nm.

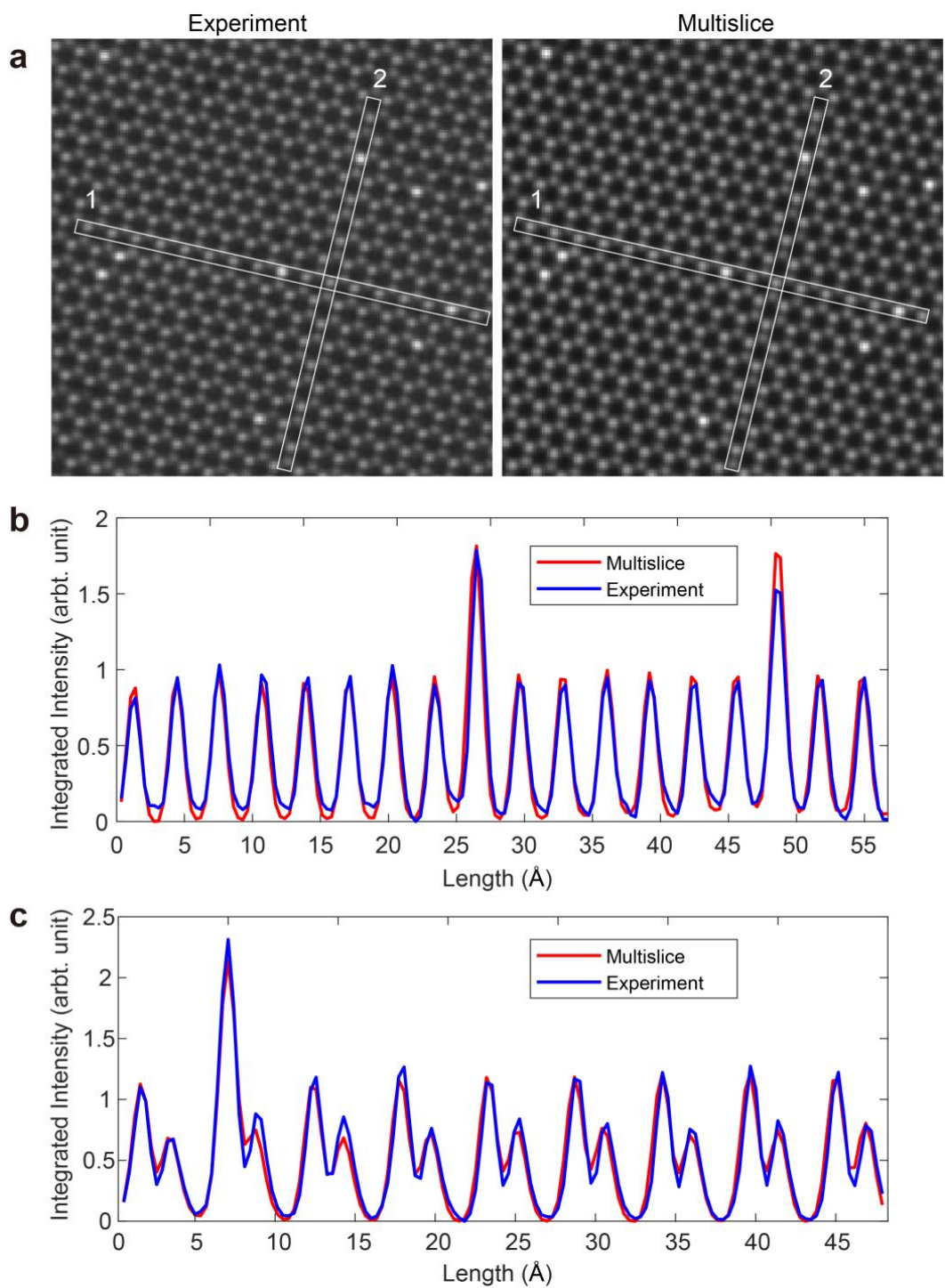

**Supplementary Figure 8 | Multislice simulation results. a,** The experiment image versus the multislice image of the same region. **b** and **c**, The intensity profiles corresponding to labeled regions 1 and 2 in (**a**), where the high, medium and low intensity peaks represent the Re, Mo and S atoms, respectively.

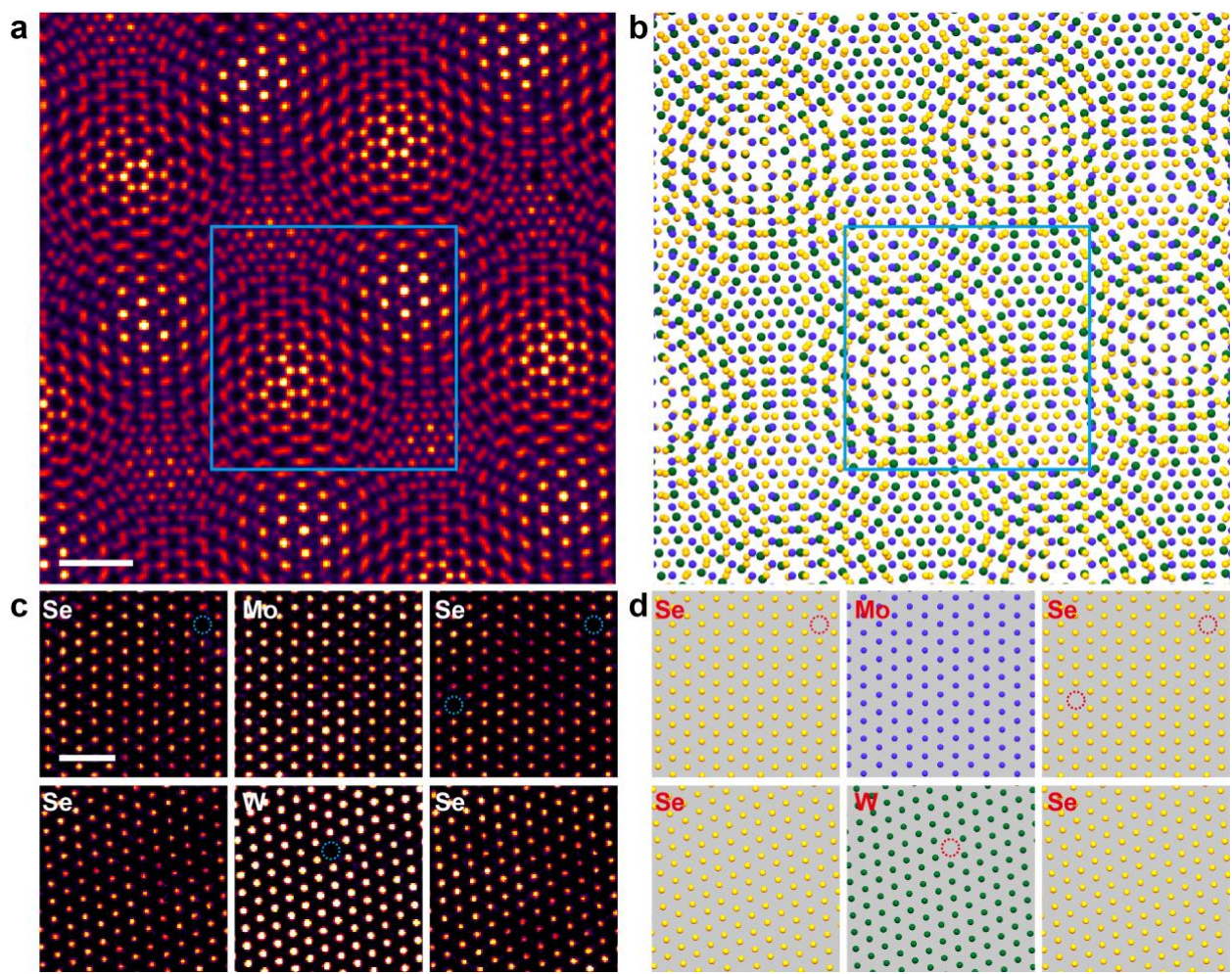

**Supplementary Figure 9 | Numerical simulations on the sAET reconstruction of a MoSe₂-WSe₂ heterostructure with moiré patterns. a,** Top view of the 3D reconstruction of double-layered heterostructure from 15 multislice projections with a double tilt range from -25° to +25°. **b,** Top view of 3D atomic model obtained from (**a**), where a 5° rotational mismatch between the top and bottom layers is visible. **c,** 6 atomic layers of the 3D reconstruction in the square region in (**a**), where atomic defects in each layer are labelled as blue circles. **d,** 6 atomic layers of the 3D model in the square region in (**b**), where traced defects are shown as red circles. The RMSD between the sAET reconstruction and the original atomic model is 8 pm, 5 pm and 13 pm for Mo, W and Se atoms, respectively.

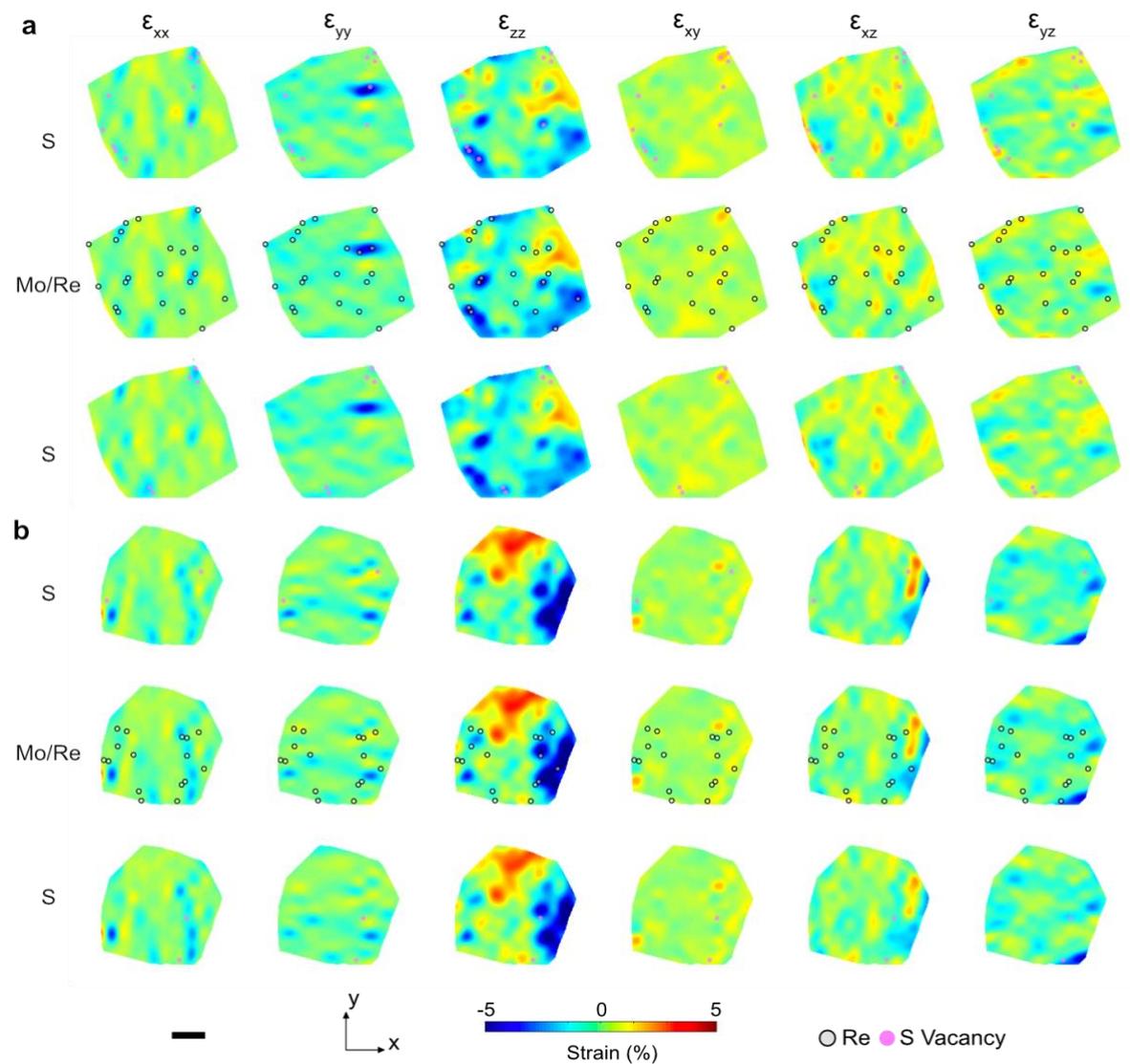

**Supplementary Figure 10 | Full 3D strain tensor of data set 1 (a) and data set 2 (b).** The six strain components are shown from left to right columns, and the three atomic layers are displayed from top to bottom. Scale bar, 2 nm.

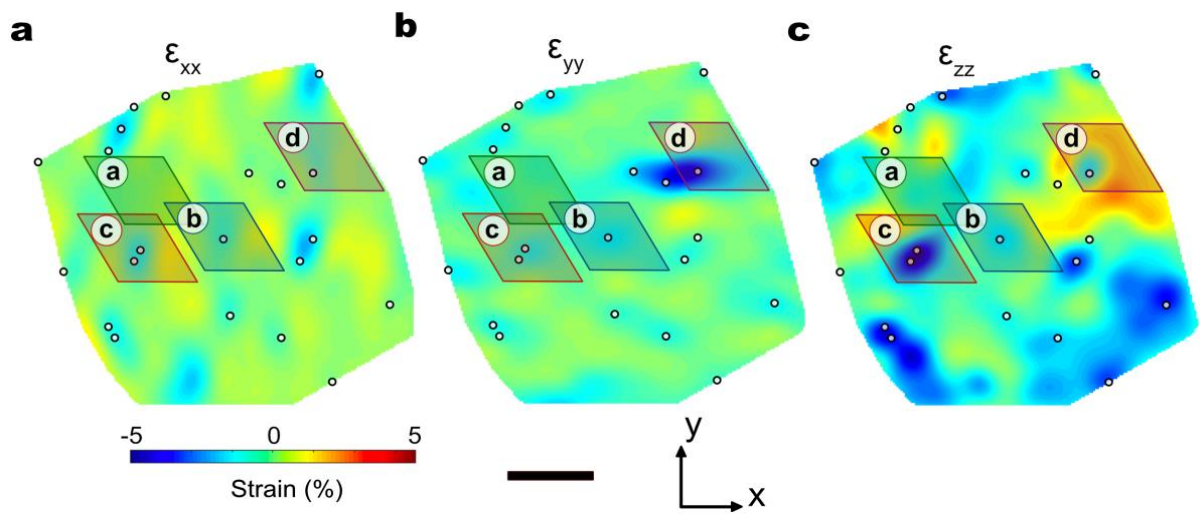

**Supplementary Figure 11 |** Strain tensor maps of $\varepsilon_{xx}$ (**a**), $\varepsilon_{yy}$ (**b**), $\varepsilon_{zz}$ (**c**) of the Mo/Re layer in data set 1 are highlighted with areas representing four 6×6×1 supercells used in DFT calculations with labels (**a**), (**b**), (**c**) and (**d**) in Fig. 5. Scale bar, 2nm.

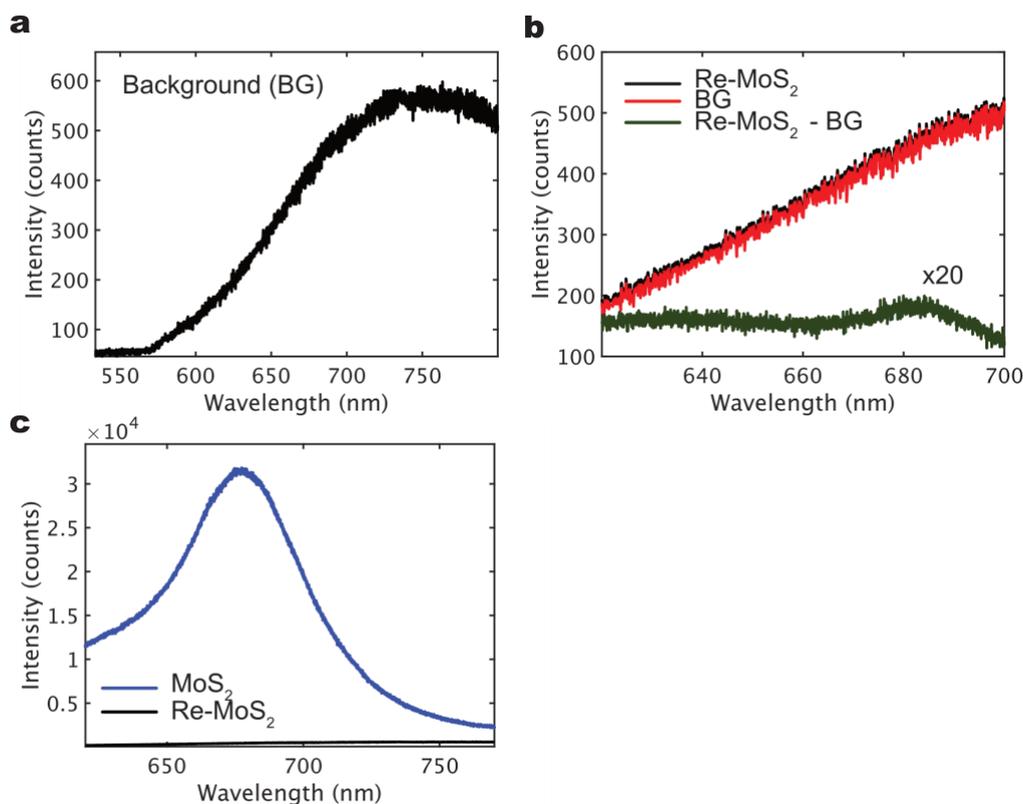

**Supplementary Figure 12 | Photoluminescence spectra measurement. a,** Si-SiO$_2$ background. **b,** Re-MoS$_2$ spectrum (black), background (red), and Re-MoS$_2$ spectrum after background subtraction (green). **c,** Overall comparison between the measurements from pristine MoS$_2$ (blue) and Re-MoS$_2$ (black)**.**

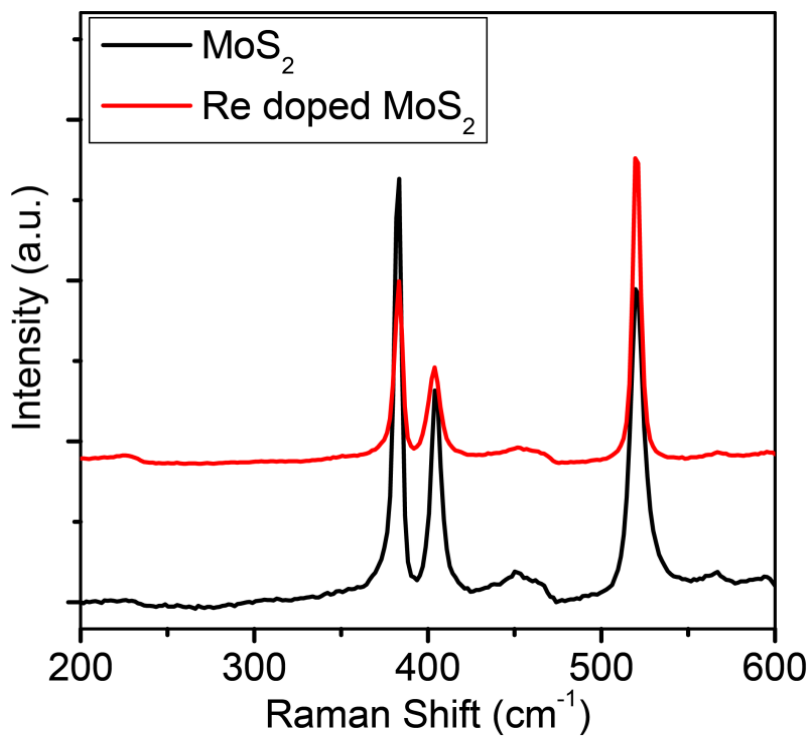

**Supplementary Figure 13 | Raman spectra measurement.** Raman spectra measured from pristine and Re-doped $MoS_2$ monolayer samples. The peak position and peak intensity of Si at 521 cm$^{-1}$ are identical for the two spectra, which is a reference for calibration and intensity comparison. For both the in-plane ($E_{2g}$) and out-of-plane ($A_{1g}$) phonon modes, the peak intensity from Re-doped $MoS_2$ (red) is about 1/3 of that from pristine $MoS_2$ (black).

**Supplementary Table 1 | sAET data collection, reconstruction, refinement statistics.**

| | Data set 1 | Data set 2 |
|---|---|---|
| **Data collection and processing** | | |
| Voltage (kV) | 60 | 60 |
| Convergence semi-angle(mrad) | 30 | 30 |
| Probe size (Å) | 1.3 | 1.3 |
| Detector inner angle (mrad) | 30 | 30 |
| Detector outer angle (mrad) | 300 | 300 |
| Depth of focus (nm) | 10 | 10 |
| Pixel size (Å) | 0.34 | 0.34 |
| # of projections | 13 | 13 |
| Tilt range | See Suppl. Table 2 | See Suppl. Table 2 |
| Electron dose ($10^5$ e/Å$^2$) | 4.1 | 4.1 |
| | | |
| **Reconstruction** | | |
| Algorithm | GENFIRE | GENFIRE |
| Interpolation radius (voxel) | 0.3 | 0.3 |
| Oversampling ratio | 3 | 3 |
| Number of iterations | 1,000 | 1,000 |
| | | |
| **Refinement** | | |
| $R_1$ (%)[a] | 13.4 | 13.0 |
| $R$ (%)[b] | 25.5 | 24.2 |
| B' factors (Å$^2$) | | |
|     Mo atoms | 13.4 | 14.6 |
|     S atoms | 13.0 | 13.0 |
|     Re atoms | 14.2 | 15.8 |
| # of total atoms | 2,103 | 1,634 |
| # of Mo atoms | 686 | 531 |
| # of S atoms | 1,381 | 1,083 |
| # of Re atoms | 21 | 16 |
| # of S vacancies | 15 | 4 |

[a]The $R_1$-factor is defined as equation (5) in ref. 23. [b]The $R$ factor was calculated by $R = \frac{\sum ||F_{obs}| - |F_{calc}||}{\sum |F_{obs}|}$, where $|F_{obs}|$ is the Fourier magnitude obtained from experimental projections and $|F_{calc}|$ is the Fourier magnitude calculated from a 3D atomic model.

**Supplementary Table 2 | Angle calibration for a double-tilt Nion stage.** The nominal angles were read out from the Nion microscope. The angles were calibrated by 3D coordinate fitting of the Re and Mo atoms using least square minimization.

| | Nominal angles (°) | | Calibrated angles (°) Data set 1 | | Calibrated angles (°) Data set 2 | |
|---|---|---|---|---|---|---|
| Tilt axis | α | β | α | β | α | β |
| | [0 -1 0] | [-1 0 0] | [-0.1 1.0 0] | [-1.0 0.2 0.2] | [-0.2 1.0 0] | [-1.0 0.2 0.2] |
| Projection #1 | 0 | 0 | 0 | 0 | 0 | 0 |
| #2 | -14.3 | -22.9 | -15.1 | -17.8 | -15.1 | -16.8 |
| #3 | -14.3 | 17.1 | -12.6 | 16.8 | -11.3 | 17.1 |
| #4 | 20.1 | 17.1 | 19.1 | 18.2 | 20.7 | 13.1 |
| #5 | 20.1 | -20.1 | 17.7 | -19.8 | 18.3 | -22.4 |
| #6 | 12.6 | -12.6 | 13.1 | -12.0 | 13.6 | -13.5 |
| #7 | 12.6 | 12.6 | 13.0 | 12.8 | 13.6 | 11.8 |
| #8 | -12.6 | 12.6 | -13.1 | 13.6 | -13.6 | 13.8 |
| #9 | -12.6 | -12.6 | -14.4 | -9.2 | -14.3 | -8.1 |
| #10 | -6.3 | -6.3 | -8.2 | -3.6 | -7.5 | -3.5 |
| #11 | -6.3 | 6.3 | -7.6 | 8.0 | -7.5 | 8.7 |
| #12 | 6.3 | 6.3 | 5.2 | 8.5 | 9.4 | 11.2 |
| #13 | 6.3 | -6.3 | 4.9 | -4.9 | 7.3 | -5.5 |

**Supplementary Video 1.** 3D sAET reconstruction of the Re-doped $MoS_2$ monolayer (data set 1), showing the different intensity distribution of the Re, Mo and S atoms as well as the S vacancies.

**Supplementary Video 2**. 3D atomic model of the Re-doped $MoS_2$ monolayer (data set 1) determined by sAET, consisting of 1380 S (in yellow), 686 Mo (in blue), 21 Re atoms (in black) and 16 S vacancies (in pink). Compared to an ideal $MoS_2$ atomic model (in lighter colors), the experimental model shows 3D crystal defects, atomic displacements and full strain tensors of the 2D material. The local strains induced by single Re dopants are also visible.